\documentclass[twocolumn,a4paper,reprint,amssymb,aps,prd,groupedaddress,superscriptaddress,nofootinbib]{revtex4-2}
\pdfoutput=1
\usepackage{booktabs}
\usepackage{multirow}
\usepackage{dcolumn}
\usepackage{bm}
\usepackage{amsmath}
\usepackage{mathtools}
\usepackage{amsfonts}
\usepackage{pbox}
\usepackage{color}
\usepackage[dvipsnames]{xcolor}
\usepackage{tabularx}
\usepackage{subfigure}
\usepackage{graphicx}
\usepackage{psfrag}
\usepackage{lipsum}
\usepackage{enumitem}
\usepackage{changes}
\usepackage{array}
\usepackage{CJK}
\usepackage[english]{babel}
\DeclareMathOperator{\arcsinh}{arcsinh}
\usepackage[colorlinks = true,
            linkcolor = blue,
            urlcolor  = blue,
            citecolor = blue,
            anchorcolor = blue]{hyperref}
\usepackage{blindtext}
\usepackage[utf8]{inputenc}
\usepackage{graphicx}
\usepackage{placeins}
\usepackage{siunitx}
\usepackage{amssymb}
\usepackage{mathtools}
\usepackage{physics}
\usepackage{tikz}
\usepackage[titletoc]{appendix}
\usepackage{listings}
\usepackage{float}
\usepackage{braket}
\usepackage{soul}
\usepackage{array}

\begin{document}

\title{Relaxing cosmological tensions with a sign switching cosmological constant}


\author{\"{O}zg\"{u}r Akarsu}
\email{akarsuo@itu.edu.tr}
\affiliation{Department of Physics, Istanbul Technical University, Maslak 34469 Istanbul, Turkey}

\author{Suresh Kumar}
\email{suresh.math@igu.ac.in}
\affiliation{Department of Mathematics, Indira Gandhi University, Meerpur, Haryana-122502, India}

\author{Emre \"{O}z\"{u}lker}
\email{ozulker17@itu.edu.tr}
\affiliation{Department of Physics, Istanbul Technical University, Maslak 34469 Istanbul, Turkey}
\affiliation{{Aix Marseille Univ, Universit\'e de Toulon, CNRS, CPT, 13000 Marseille, France}}

\author{J. Alberto Vazquez}
\email{javazquez@icf.unam.mx}
\affiliation{Instituto de Ciencias F\'isicas, Universidad Nacional Aut\'onoma de M\'exico, Cuernavaca, Morelos, 62210, M\'exico}

\begin{abstract}
Inspired by the recent conjecture originated from graduated dark energy that the Universe has recently transitioned from anti-de Sitter vacua to de Sitter vacua, we extend the standard $\Lambda$CDM model by a cosmological constant ($\Lambda_{\rm s}$) that switches sign at a certain redshift $z_\dagger$, and we call this model $\Lambda_{\rm s}$CDM. We discuss the construction and theoretical features of this model in detail and find out that, when the consistency of the $\Lambda_{\rm s}$CDM model with the cosmic microwave background (CMB) data is ensured, (i) $z_\dagger\gtrsim1.1$ is implied by the condition that the Universe monotonically expands, (ii) $H_0$ and $M_B$ (type Ia supernovae absolute magnitude) values are inversely correlated with $z_\dagger$ and reach $H_0\approx74.5~{\rm km\, s^{-1}\, Mpc^{-1}}$ and $M_B\approx-19.2\,{\rm mag}$ for $z_\dagger=1.5$, in agreement with the SH0ES measurements, and (iii) $H(z)$ presents an excellent fit to the Ly-$\alpha$ measurements provided that $z_\dagger\lesssim 2.34$. We further investigate the model constraints by using the full \textit{Planck} CMB data set, with and without baryon acoustic oscillation (BAO) data. We find that the CMB data alone does not constrain $z_\dagger$, but the CMB+BAO data set favors the sign switch of $\Lambda_{\rm s}$ providing the constraint: $z_\dagger=2.44\pm0.29$ (68\% C.L.). Our analysis reveals that the lower and upper limits of $z_\dagger$ are controlled by the Galaxy and Ly-$\alpha$ BAO measurements, respectively, and the larger $z_{\dagger}$ values imposed by the Galaxy BAO data prevent the model from achieving the highest local $H_0$ measurements. In general, the $\Lambda_{\rm s}$CDM model (i) relaxes the $H_0$ tension while being fully consistent with the tip of the red giant branch measurements, (ii) relaxes the $M_B$ tension, (iii) removes the discrepancy with the Ly-$\alpha$ measurements, (iv) relaxes the $S_8$ tension, and (v) finds a better agreement with the big bang nucleosynthesis constraints on the physical baryon density. We find no strong statistical evidence to discriminate between the $\Lambda_{\rm s}$CDM and $\Lambda$CDM models. However, interesting and promising features of the $\Lambda_{\rm s}$CDM model, which we describe in our study, provide an advantage over $\Lambda$CDM.
\end{abstract}
\maketitle

\section{Introduction}
\label{sec:intro} 
Over the last few years, there has been a growing consensus that the standard cosmological model---the so-called Lambda cold dark matter ($\Lambda$CDM) model---could in fact be an approximation to a more realistic one that still needs to be fully understood \cite{DiValentino:2020vhf}. Phenomenologically, this new model is not expected to deviate drastically from $\Lambda$CDM, which is in excellent agreement with most of the currently available data \cite{Riess:1998cb,Ade:2015xua,Alam:2016hwk,Abbott:2017wau,Planck:2018vyg}; however, it could be conceptually very different, and its deviations could be nontrivial. The recent developments, both theoretical (e.g., the de Sitter swampland conjecture \cite{Obied:2018sgi,Agrawal:2018own,Colgain:2018wgk,Heisenberg:2018yae,Akrami:2018ylq,Raveri:2018ddi,Cicoli:2018kdo,Colgain:2019joh}) and observational (e.g., the tensions hint at some unexpected and/or nontrivial deviations from $\Lambda$CDM; see Refs.~\cite{Freedman:2017yms,Verde:2019ivm,DiValentino:2017gzb,Tamayo:2019gqj,Aubourg:2014yra,Zhao:2017cud,Bullock:2017xww,tension02,Mortsell:2018mfj,Dutta:2018vmq,Hee:2016ce,DiValentino:2017rcr,Vagnozzi:2019ezj,Handley:2019tkm,DiValentino:2019qzk,Akarsu:2019hmw,DiValentino:2020hov,Vagnozzi:2020zrh,Delubac:2014aqe,Dutta:2018vmq,Sahni:2014ooa,Poulin:2018zxs,Capozziello:2018jya,Wang:2018fng,Banihashemi:2018oxo,Banihashemi:2018has,Farhang:2020sij,Banihashemi:2020wtb,Visinelli:2019qqu,kunuya2019,Ye:2020btb,Ye:2020oix,Perez:2020cwa,Bonilla:2020wbn,kumar:2021,araujo:2021,Vazquez:2020ani,Vagnozzi:2018jhn,DiValentino:2019exe,DiValentino:2019dzu,Calderon:2020hoc,Vazquez:2012ag,Paliathanasis:2020sfe,Akarsu:2020yqa,Efstathiou:2020wem,Vagnozzi:2020dfn,Banerjee:2020xcn,Acquaviva:2021gty,LinaresCedeno:2021aqk,Zhou:2021xov,Adil:2021zxp,DeFelice:2020sdq,LinaresCedeno:2020uxx}, and Refs.~\cite{DiValentino:2020zio,DiValentino:2020vvd,DiValentino:2020srs,Perivolaropoulos:2021jda} for more references), along with the cosmological constant problems \cite{Weinberg:1988cp,Peebles:2002gy}, suggest that attaining it would be an elusive task. These tensions are of great interest, not only in cosmology, but also in theoretical physics, as they could imply new physics beyond the well established fundamental theories that underpin, and even extend, the $\Lambda$CDM model. The so-called $H_{0}$ tension---the deficit in the Hubble constant ($H_0$) predicted by the \textit{Planck} cosmic microwave background (CMB) data within the $\Lambda$CDM model \cite{Planck:2018vyg} when compared to its model-independent determinations from local measurements of distances and redshifts \cite{Riess:2016jrr,Riess:2018byc,Riess:2019cxk,Freedman:2019jwv,Yuan:2019npk,Freedman:2020dne,Riess:2020fzl}---among others, is now described by many as a crisis. See Ref.~\cite{DiValentino:2020zio} for a comprehensive list of references on the $H_0$ tension, and Ref.~\cite{DiValentino:2021izs} for a recent comprehensive review, including a discussion of recent $H_0$ estimates and a summary of the proposed theoretical solutions. It has turned out to be a more challenging problem than originally thought as it worsens when the cosmological constant ($\Lambda$) is replaced by generic quintessence models of dark energy (DE), and is only partially relaxed when replaced by the simplest phantom (or quintom) models \cite{Vagnozzi:2018jhn,DiValentino:2019exe,DiValentino:2019dzu,Vazquez:2020ani,Banerjee:2020xcn}. Notably, it was reported that the $H_{0}$ tension---as well as a number of other low-redshift discrepancies---could be alleviated by a dynamical DE that assumes negative or rapidly vanishing energy density values at high redshifts \cite{Delubac:2014aqe,Aubourg:2014yra,Sahni:2014ooa,DiValentino:2017rcr,Mortsell:2018mfj,Poulin:2018zxs,Capozziello:2018jya,Wang:2018fng,Dutta:2018vmq,Banihashemi:2018oxo,Banihashemi:2018has,Farhang:2020sij,Banihashemi:2020wtb,Visinelli:2019qqu,Akarsu:2019hmw,Ye:2020btb,Ye:2020oix,Perez:2020cwa,Calderon:2020hoc,Paliathanasis:2020sfe,Bonilla:2020wbn,Vazquez:2012ag,Akarsu:2020yqa,LinaresCedeno:2021aqk,Zhou:2021xov,LinaresCedeno:2020uxx}. The fact that the \textit{Planck} CMB data alone favors positive spatial curvature ($\Omega_{k0}<0$), on top of the $\Lambda$CDM model, suggests that curvature might be the simplest explanation for a negative energy density source (effectively); however, the drastic exacerbation of the $H_0$ tension for the $\Lambda$CDM model with spatial curvature, and the favoring of spatial flatness ($\Omega_{k0}=0$) with extremely high precision by the \textit{Planck} CMB data in combination with other astrophysical data such as baryon acoustic oscillations (BAO) and cosmic chronometers, indicate that the negative energy source cannot be spatial curvature, but a nontrivially evolving DE \cite{Planck:2018vyg,Handley:2019tkm,DiValentino:2019qzk,DiValentino:2020hov,Vagnozzi:2020zrh,Efstathiou:2020wem,Vagnozzi:2020dfn,Acquaviva:2021gty}.
 
The CMB power spectrum by itself, for a given cosmological model, provides powerful constraints on the Hubble parameter $H(z)$ at the background level once the comoving sound horizon at CMB last scattering, $r_*$, is given \cite{Dodelson03,Knox:2019rjx,DiValentino:2021izs}. The comoving sound horizon at last scattering is determined entirely by the pre-recombination Universe, and is given by $r_{*}=\int^{\infty}_{z_{*}}c_{\rm s}H^{-1}\dd{z}$, where $c_{\rm s}$ is the sound speed in the plasma and $z_{*}\approx 1100$ is the redshift of last scattering.
The acoustic angular scale on the sky, $\theta_*$, which is measured almost model independently with a precision of 0.03\% \cite{Planck:2018vyg}, determines the comoving angular diameter distance to last scattering $D_M(z_*)$ through the relation $D_M(z_*)=r_*/\theta_*$. The measured CMB monopole temperature determines the radiation energy density, and the positions and heights of the angular peaks determine $\rho_{\rm c}(z_*)$ and $\rho_{\rm b}(z_*)$, where $\rho$ is the energy density and the indices stand for CDM and baryonic matter, respectively. Assuming a flat space, $D_M(z_{*})=c\int_0^{z_{*}}H^{-1}\dd{z}$, where $c$ is the speed of light (unless it is mentioned explicitly, we will use $c=1$); for $\Lambda$CDM, the constraints from the CMB along with this integral are enough to infer the value of $\Lambda$ and hence the complete evolution of $H(z)$. These steps make it clear how phantom/quintom extensions of $\Lambda$CDM, for which $\Lambda$ is replaced by a DE density typically decreasing and approaching to zero with increasing redshift, increase $H_0$. The decreased DE density at high redshifts corresponds to a lower $H(z)$ at those redshifts compared to $\Lambda$CDM. Since $D_M(z_*)$ is the same to very high precision for different DE models, the decreased $H(z)$ at higher redshifts should be compensated by an increased $H(z)$ at lower redshifts (and hence an increased $H_0$) in order to keep the integral describing $D_M(z_*)$ unaltered. This also explains why quintessence models exacerbate the $H_0$ tension: these models have a DE density that increases with redshift, so the above mechanism is reversed. Note that, the DE density is negligible in these models for $z>z_*$ as in $\Lambda$CDM, so $r_*$ is not affected by the dynamical nature of the DE. Nevertheless, the simplest phantom/quintom models can only partially relieve the $H_0$ tension \cite{Vagnozzi:2018jhn,DiValentino:2019exe,DiValentino:2019dzu,Vazquez:2020ani,Banerjee:2020xcn}; however, a DE density that attains negative values at high redshifts can amplify this mechanism to enhance $H_0$ even further. We recall that the above discussion relies on $r_*$ being fixed among different models, in contrast to models that modify the sound horizon to alter $D_M(z_*)$ and hence $H_0$, e.g., early dark energy (EDE) models \cite{Poulin:2018cxd}.

On top of increasing $H(z)$ at low redshifts and hence the $H_0$ value, a lower $H(z)$ at large redshifts compared to the $\Lambda$CDM model can provide better agreement with the Ly-$\alpha$ BAO measurements at the effective redshift $z\sim2.34$ \cite{Agathe:2019vsu,Blomqvist:2019rah}, if the drop in the DE density is large enough at that redshift. Also, if the drop is rapid enough, it can cause a nonmonotonic behavior of $H(z)$ which is hard to achieve without relying on a negative DE density. Such a nonmonotonic behavior can provide an even better description of the Ly-$\alpha$ data, and was initially suggested by the BOSS Collaboration after the BOSS DR11 data  \cite{Delubac:2014aqe} presented an approximately $2.5\sigma$ discrepancy with the best-fit $\Lambda$CDM model of \textit{Planck} 2015 \cite{Ade:2015xua}. They have also reported, in a companion paper \cite{Aubourg:2014yra}, that a positive cosmological constant is consistent with their data set for $z<1$, while a negative DE density is preferred for $z>1.6$, which led them to suggest a nonmonotonic behavior of $H(z)$ at $z\sim2$. The \textit{Planck} Collaboration (2018)  \cite{Planck:2018vyg} does not include the Ly-$\alpha$ measurements in their default BAO data compilation since for the $\Lambda$CDM model and its simple extensions, they do not provide significant constraints once the CMB and Galaxy BAO data are used, and they do not conform well with the rest of the data set within the framework of these models. They also quote from  \cite{Delubac:2014aqe} that well-motivated extensions of $\Lambda$CDM that could provide a resolution to this discrepancy are hard to construct. Currently, the discrepancy of the Ly-$\alpha$ measurements with the \textit{Planck} 2015 best-fit $\Lambda$CDM is reduced to a mild $\sim1.7\sigma$ when the combination of the BOSS survey and its extended version eBOSS in the SDSS DR14 \cite{Agathe:2019vsu,Blomqvist:2019rah} is considered, and reduced even further to a $\sim1.5\sigma$ tension when the final eBOSS (SDSS DR16) measurement, which combines all the data from eBOSS and BOSS \cite{Alam:2020sor,duMasdesBourboux:2020pck}, is considered. We note, however, that since $H_0$ values predicted by $\Lambda$CDM
are lower than the local measurements of $H_0$ while $H(z)$ values predicted by $\Lambda$CDM at $z\sim2.34$ are greater than the Ly-$\alpha$ measurements of $H(z)$, simple and/or well-motivated extensions of $\Lambda$CDM addressing either one of these discrepancies typically tend to exacerbate the other. Therefore, it is conceivable that such models relaxing the $H_0$ tension will also typically suffer from a greater tension with the Ly-$\alpha$ measurements. It is intriguing to note that the Ly-$\alpha$ discrepancy has certain parallelisms with the so-called $S_8$ discrepancy (quantifying a discordance between the CMB and low redshift probes, and will be further elaborated in Sec. \ref{sec:obs}), e.g., $S_8$ constraints based on Ly-$\alpha$ measurements are in agreement with the low redshift probes \cite{Palanque-Delabrouille:2019iyz}, simple extensions of $\Lambda$CDM that reduce the $H_0$ tension typically worsen the $S_8$ discrepancy and vice versa \cite{DiValentino:2020vvd}, and the $S_8$ discrepancy has also weakened with the latest observations \cite{Hamana:2019etx,vanUitert:2017ieu}. These facts seem to hint that a model addressing the $H_0$ and Ly-$\alpha$ tensions simultaneously may also address the $S_8$ tension. With all of these in hand, a DE density that is consistent with a positive cosmological constant today but assumes negative values in the past is not indispensable, and yet it is worth further investigation as it has the potential to result in a better agreement with the existing observational data, including Ly-$\alpha$, while addressing the $H_0$ tension too.

In this paper, we study a simple extension of the $\Lambda$CDM model for which a cosmological constant that yields a negative value in the past switches sign at certain redshift $z_\dagger$ to attain its current positive value and drives the observed acceleration; it will be dubbed $\Lambda_{\rm s}$CDM. Although this sign switch results in discontinuities in various fundamental functions, e.g., in $H(z)$, it can be considered as an approximation to a rapid transition in the (possibly effective) DE density. In fact, the sign switching feature of the $\Lambda_{\rm s}$CDM model was first suggested in Ref.~\cite{Akarsu:2019hmw} when their graduated dark energy (gDE) model appeared to prefer a very rapid transition in the DE density resembling a step function whose absolute value is almost constant away from the transition point. In Sec. \ref{sec:cosmo}, we first we motivate the $\Lambda_{\rm s}$CDM model starting from the gDE, and then study its theoretical features. In Sec. \ref{sec:obs}, we conduct a robust observational analysis of the model with the latest data, and, we conclude in Sec. \ref{sec:conc}.

\section{\texorpdfstring{$\bm{\Lambda_{\rm s}}$}{$\Lambda_{\rm s}$}CDM model: Sign-switching \texorpdfstring{$\bm{\Lambda}$}{$\Lambda$}}
\label{sec:cosmo}

The positive cosmological constant assumption of the $\Lambda$CDM model was investigated via the gDE characterised by a minimal dynamical deviation from the null inertial mass density $\varrho=0$ (where $\varrho\equiv\rho+p$) of the cosmological constant---or, the usual vacuum energy of the quantum field theory (QFT). This deviation is in the form of $\varrho\propto \rho^{\lambda}<0$, for which, provided that the parameter $\lambda<1$ is the ratio of two odd integers, the energy density $\rho$ dynamically takes negative values in the past \cite{Akarsu:2019hmw}. During the transition from negative to positive energy density, there comes a redshift for which the energy density is null; this redshift will be denoted by $z_\dagger$ in the present work, but note that it was denoted by $z_*$ in Ref.~\cite{Akarsu:2019hmw}. gDE exhibits a wide variety of behaviors depending on $\lambda$, but it is of particular interest to us that for large negative values of $\lambda$, it establishes a phenomenological model characterized by a smooth function that approximately describes a $\Lambda$ that switches sign in the late Universe to become positive today. It was shown via the gDE that the joint observational data, including but not limited to the \textit{Planck} CMB and Ly-$\alpha$ BAO (BOSS DR11) data, suggest that the cosmological constant changed its sign at $z\approx 2.32$ and triggered the late-time acceleration, the behavior of which alleviates the $H_0$ tension and the discrepancy with the Ly-$\alpha$ BAO measurements simultaneously. For large negative values of $\lambda$, it turns out that $\rho_{\rm gDE}/3H_0^2\approx0.70$ for $0\leq z\lesssim2.32$, but its energy density switches sign rapidly at $z_\dagger\approx 2.32$ (this $z_\dagger$ value is quite stable for $\lambda\lesssim-4$) and settles into a value $\rho_{\rm gDE}/3H_0^2\sim-0.70$ and remains there for $z_\dagger\gtrsim 2.32$; moreover, the larger the negative values of $\lambda$, the more $\rho_{\rm gDE}$ resembles a step function, and the better fit to the data. For arbitrarily large negative values of $\lambda$, $\rho_{\rm gDE}$ indeed transforms into a step function centred at $z_\dagger$ with two branches yielding opposite values about zero. It is easy to check that $\lambda$ is responsible from the rapidity of the sign change of the energy density, and for the constraint $\lambda=-17.9\pm5.8$ obtained on it, the function $\rho_{\rm gDE}(z)$ already closely resembles a step function. Thus, the gDE suggesting large negative values of $\lambda$ when confronted with the observations can be interpreted as a hint at a cosmological constant that achieved its present-day positive value by switching sign at $z_\dagger\sim2.3$, but was negative in the earlier Universe. 
 
 Some general constraints that are typically applied to classical sources, irrespective of a detailed description, give further confidence to the interpretation of the gDE as a hint at a sign-switching cosmological constant \cite{EllisRC,Carroll:2003st}. Let us consider the gDE as an actual barotropic fluid, $p=p(\rho)$. In this case, although it behaves almost like a cosmological constant (in spite of the fact that its value switches sign at $z\approx2.32$) throughout the history of the Universe, strictly speaking, it violates the weak energy condition, namely, the non-negativity conditions on the energy density, $\rho\geq0$, for $z>z_\dagger$, and on the inertial mass density, $\varrho\geq0$, at any given time. Moreover, there are phases during which $c_{\rm s}^2\gg1$ and $c_{\rm s}^2<0$, i.e., gDE violates the condition $0\leq c_{\rm s}^2\leq1$ on the speed of sound of a barotropic fluid given by the adiabatic formula $c_{\rm s}^2={\rm d}p/{\rm d}\rho$. The upper limit (causality limit) is a rigorous limit, and its violation means the abandonment of the theory of relativity. The lower limit applies to a stable situation, and if violated, the fluid is classically unstable against small perturbations of its background energy density---the so-called Laplacian (or gradient) instability. Indeed, phenomenological fluid models of DE are difficult to motivate, and adiabatic fluid models are typically unstable against perturbations, since $c_{\rm s}^2$ is usually negative for $w= p/\rho<0$. It is possible to evade this constraint in adiabatic fluids---such as canonical scalar field (quintessence or phantom fields) and string-theory-inspired tachyon fields, for which the effective speed of sound $c_{\rm s\, eff}$ (which governs the growth of inhomogeneities in the fluid) remains consistent with $0\leq c_{\rm s\, eff}^2\leq1$---in adiabatic fluids if $w$ decreases sufficiently fast as the Universe expands (e.g., Chaplygin gas), and in multi-fluid models of DE (e.g., quintom field) constructed from the combination of such fluids \cite{Copeland:2006wr}. However, unlike such sources, it seems unlikely to evade this constraint in gDE, especially given the observationally preferred values of its free parameters. On the other hand, whether it is positive or negative, a cosmological constant, which corresponds to the $\lambda\rightarrow-\infty$ limit of the gDE, is well behaved: $\varrho=0$, and $c_{\rm s}^2=0$ (it has no speed of sound, and thereby does not support classical fluctuations). Regarding the negativity of the corresponding energy density (when $z>z_\dagger$), a negative cosmological constant is not only ubiquitous in the fundamental theoretical physics without any complication, but also a theoretical sweet spot; an anti-de Sitter (AdS) background (provided by $\Lambda<0$) is welcome due to the celebrated AdS/CFT (conformal field theory) correspondence \cite{Maldacena:1997re} and is preferred by string theory and string-theory-motivated supergravities \cite{Bousso:2000xa}. It is the positive cosmological constant that in fact suffers from theoretical challenges: getting a vacuum solution with a positive cosmological constant within string theory or formulating QFT on the background of a dS space (provided by $\Lambda>0$) has been a notoriously difficult task [see Refs.~\cite{Obied:2018sgi,Ooguri:2006in,Maldacena:2000mw,Kachru:2003aw,Conlon:2007gk,Danielsson:2018ztv,Witten:2001kn,Goheer:2002vf}; additionally, see Refs.~\cite{Cicoli:2018kdo,Dutta:2021bih} for a recent review on models of the accelerating Universe (viz., for different mechanisms to obtain dS space/vacua and building models of quintessence) in supergravity and string theory]. Therefore, an approach that asserts that a positive-valued cosmological constant exists only in the late Universe (say, when $z\lesssim2.3$) would enjoy limiting such difficulties to the late Universe. Of course, it is necessary to further study whether such an approach---say, transitions from AdS background to dS one---would be viable both theoretically and observationally (we further comment on such transitions in Sec. \ref{sec:conc}). Besides, studies considering the presence of a negative cosmological constant in various contexts are already plentiful in the cosmology literature. In the context of inflationary Universe, see, e.g., Refs.~\cite{Piao:2004me,Li:2019ipk, Vazquez:2018qdg} which considered inflation with multiple AdS vacua, and Ref.~\cite{Yin:2021uus} which considered a cosmological constant that slowly varies from a positive value to a negative value and becomes vanishingly small value at the end of inflation. In the context of EDE models, see, e.g, Ref.~\cite{Ye:2020btb} which suggested the presence of AdS vacua around recombination to alleviate the $H_0$ tension, and the a follow up study in Ref.~\cite{Ye:2020oix} which presented an $\alpha$-attractor AdS model of EDE for which the AdS vacua originate from UV-complete theories in the cosmological setup with varying AdS depth. In the context of post-recombination modifications to the $\Lambda$CDM model, see, e.g., Refs.~\cite{Sahni:2014ooa,Poulin:2018zxs,Wang:2018fng,Dutta:2018vmq,Mortsell:2018mfj,Akarsu:2019hmw,Bonilla:2020wbn,Calderon:2020hoc,LinaresCedeno:2021aqk} which suggested that the cosmological data prefer or are fully consistent with the presence of a negative-valued cosmological constant at high redshifts; some of these works explicitly pronounce the redshift scales $z\gtrsim 2.3$. Let us also mention that a negative (but not necessarily constant) effective energy component appears and find applications in the cosmology literature [see, e.g., scalar-tensor theories of gravity such as Brans-Dicke theory \cite{Faraoni:1998qx,Boisseau:2000pr,Sahni:2006pa,Akarsu:2019pvi,SolaPeracaula:2020vpg}, as well as modified theories of gravity such as $f(R, \mathcal{L}_{\rm m})$ \cite{Harko:2010mv}, $f(R,T)$ \cite{Harko:2011kv}, $f(R,T_{\mu\nu}T^{\mu\nu})$ \cite{Katirci:2014sti,Roshan:2016mbt,Akarsu:2017ohj,Board:2017ign,Akarsu:2018aro,Akarsu:2019ygx}, Rastall gravity \cite{Akarsu:2020yqa}, quadratic bimetric gravity \cite{Mortsell:2018mfj}; theories in which $\Lambda$ relaxes from a large initial value via an adjustment mechanism \cite{Dolgov:1982qqA,Dolgov:1982qqB,Bauer:2010wj}; cosmological models based on Gauss-Bonnet gravity \cite{Zhou:2009cy}; braneworld models \cite{Sahni:2002dx,Brax:2003fv}; higher dimensional cosmologies that accommodate dynamical reduction of the internal space \cite{Chodos:1979vk,Dereli:1982ar,Akarsu:2012am,Akarsu:2012vv,Russo:2018akp}; a negative dark radiation component \cite{Boisseau:2015hqa}; missing matter \cite{Hee:2016ce}; a dynamical $\Lambda(t)$ term \cite{Grande:2006nn}; phenomenological generalizations of the null inertial mass density of the usual vacuum energy \cite{Bouhmadi-Lopez:2014cca,Acquaviva:2021gty,Stefancic:2004kb,Barrow:1990vx,Akarsu:2019hmw}; a negative matter action \cite{Petit:2014ura,Farnes:2017gbf,Najera:2021tcx}; and ghost-matter cosmologies \cite{Chavda:2020tfh}].
 
 Thus, bringing all of these points together, it is tempting to consider the possibility that the cosmological constant switched sign and became positive in the late Universe, which then eventually started the acceleration. Accordingly, we introduce the $\Lambda_{\rm s}$CDM model phenomenologically, constructed simply by replacing the usual cosmological constant ($\Lambda$) of the standard $\Lambda$CDM model with a cosmological constant ($\Lambda_{\rm s}$) that switches its sign from negative to positive when the Universe reaches a certain energy scale (redshift $z_\dagger$) during its expansion;
 \begin{equation}
     \Lambda\quad\rightarrow\quad\Lambda_{\rm s}\equiv\Lambda_{\rm s0}\,{\rm sgn}[z_\dagger-z],
 \end{equation}
 where $\Lambda_{\rm s0}>0$. Here ``${\rm sgn}$" is the signum function that reads ${\rm sgn}[x]=-1,0,1$ for $x<0$, $x=0$ and $x>0$, respectively. Accordingly, the Friedmann equation for the $\Lambda_{\rm s}$CDM model reads:
\begin{equation}
    \frac{H^2}{H_0^2}=\Omega_{\rm r0}(1+z)^{4}+\Omega_{\rm m0}(1+z)^{3}+\Omega_{\rm \Lambda_{\rm s}0}{\rm sgn}[z_\dagger-z],
    \label{nlcdm}
\end{equation}
where we consider the usual cosmological fluids [CDM (c) and baryons (b) described by the equation of states $w_{\rm c}=w_{\rm b}=0$, and radiation (r), consisting of photons ($\gamma$) and neutrinos ($\nu$), described by $w_{\rm r}=\frac{1}{3}$] and $\Omega_{\rm m0}+\Omega_{\rm r0}+\Omega_{\rm \Lambda_{\rm s}0}=1$ with $\Omega_{\rm m0}=\Omega_{\rm c0}+\Omega_{\rm b0}$. We define the present-day density parameters as $\Omega_{\rm r0}=8\pi G \rho_{\rm r0}/(3 H_0^2)$, $\Omega_{\rm m0}=8\pi G \rho_{\rm m0}/(3 H_0^2)$, and $\Omega_{\Lambda_{\rm s}0}=\Lambda_{\rm s0}/(3 H_0^2)$. Note that the index $0$ stands for the present-day values, but we will drop it from the indices of the density parameters in the next section to avoid cluttered notation. Accordingly, the corresponding energy density and pressure for the dark energy read $\rho_{\rm DE}=\Lambda_{\rm s0}{\rm sgn}[z_\dagger-z]/(8 \pi G)$ and $p_{\rm DE}=-\Lambda_{\rm s0}{\rm sgn}[z_\dagger-z]/(8 \pi G)$, respectively, satisfying the equation of state $p_{\rm DE}=-\rho_{\rm DE}$ like the usual vacuum energy.\footnote{Note that the signum function implies $p_{\rm DE}(z_\dagger)=-\rho_{\rm DE}(z_\dagger)=0$; however, this is an artifact of using the signum function to describe the sign switch, and is not fundamental to the model. We could, instead, make use of, e.g., the Heaviside step function which is devoid of this artifact, but this would make no meaningful contribution to our discussions, and would crowd the equations; for this reason, we stick with the familiar signum function. Furthermore, $\Lambda_{\rm s}$CDM can also be extended by modeling the sign switch with smooth sigmoid functions which would allow one to study also the rapidity of the transition, but we leave this possibility to future works.} The radiation density parameter today is given by $\Omega_{\rm r0}=2.469\times10^{-5}h^{-2}(1+0.2271N_{\rm eff})$---where $h=H_0/100\, {\rm km\,s}^{-1}{\rm Mpc}^{-1}$ is the dimensionless reduced Hubble constant and $N_{\rm eff}=3.046$ is the standard number of effective neutrino species with minimum allowed mass $m_{\nu}=0.06$ eV---as the present-day photon energy density is already extremely well constrained by the absolute CMB monopole temperature measured by FIRAS $T_0 =2.7255 \pm 0.0006 \,{\rm K}$ \cite{Fixsen09}.

\begin{figure}[t!]
    \centering
    \includegraphics[width=0.48\textwidth]{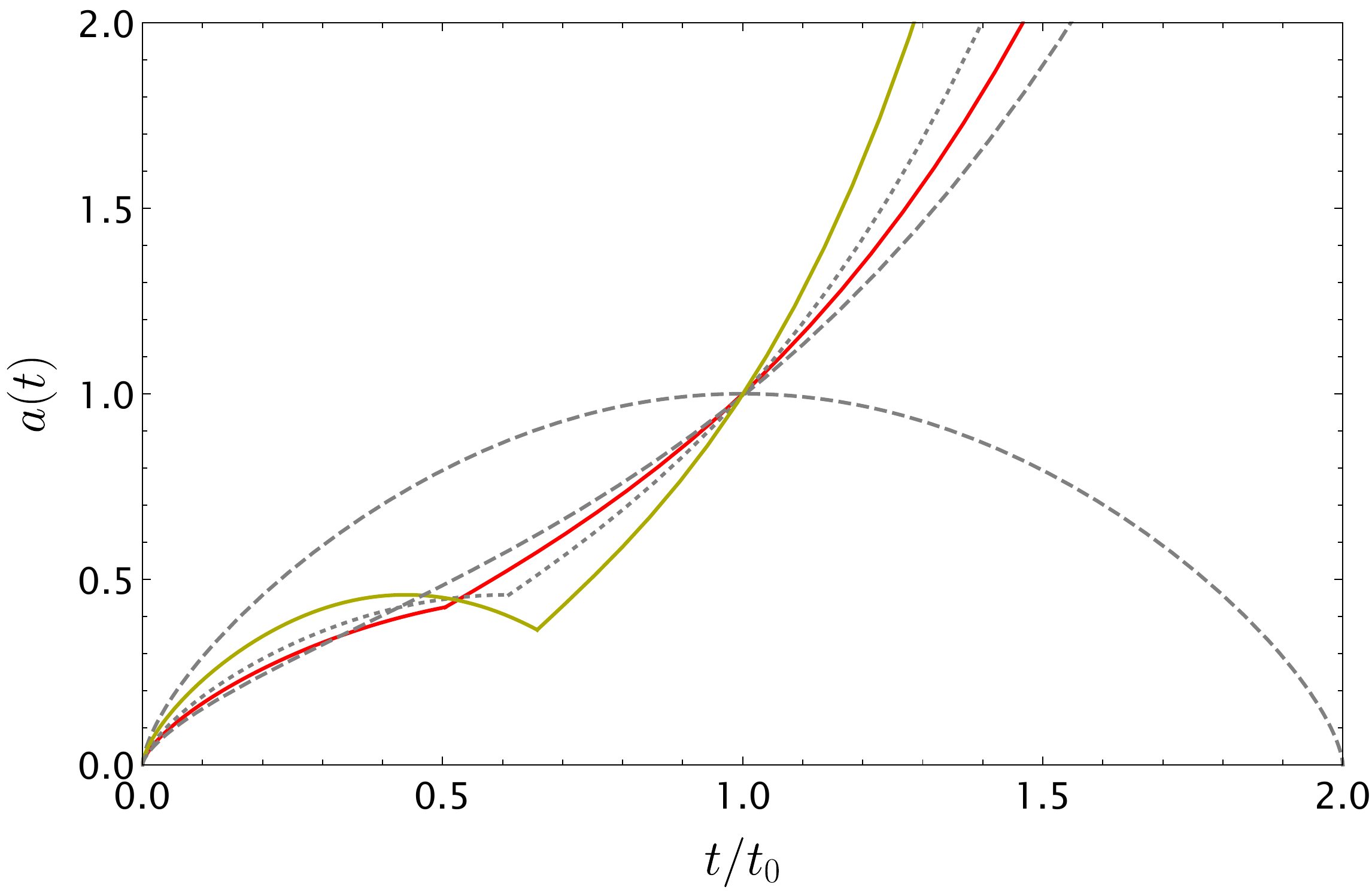}
    \caption{Evolution of the scale factor for various scenarios under the constraints $a(0)=0$ and $a(t_0)=1$. The dashed gray curves are the edge cases $t_\dagger=0$ and $t_\dagger\to\infty$, i.e., the standard Friedmann-Lema\^{i}tre models for a positive cosmological constant (which expands forever), and for a negative cosmological constant (which recollapses), respectively. The red curve corresponds to an ever-expanding Universe, i.e., $t_\dagger<\pi/\sqrt{3\Lambda_{\rm s0}}$, and is the most relevant case for this paper. The dark yellow curve is for $t_\dagger>\pi/\sqrt{3\Lambda_{\rm s0}}$, and the dotted gray curve is the critical case $t_\dagger=\pi/\sqrt{3\Lambda_{\rm s0}}$. Note that radiation is neglected in the figure, but since $t_{\rm eq}/t_0\approx0$ and $a(t_{\rm eq})\approx0$, its inclusion would not result in visible changes.}
    \label{fig:scalefac}
\end{figure}

To better understand the behavior of the $\Lambda_{\rm s}$CDM model described by the Friedmann equation in Eq. \eqref{nlcdm}, we proceed with giving the evolution of the scale factor in cosmic (proper) time $t$, i.e., $a(t)$, under the assumption that while the cosmological constant is positive ($\Lambda_{\rm s}>0$) the Universe always expands.\footnote{In the case where the Universe starts contracting before the cosmological constant switches sign to become positive, one naturally expects the positive cosmological constant to cause an expansion after the switch; however, the resumption of the contraction after the sign switch is a mathematically viable alternative that we do not investigate in this paper due to the clear evidence in favor of the present-day expansion.} When radiation dominates the Friedmann equation \eqref{nlcdm}, i.e., at the redshifts larger than the matter-radiation equality, $z>z_{\rm eq}$, like $\Lambda$CDM, $\Lambda_{\rm s}$CDM  is also well described by the Tolman model, viz., $a(t)\propto t^\frac{1}{2}$. On the other hand, when the radiation is negligible, i.e., for $z>z_{\rm eq}$, like $\Lambda$CDM, $\Lambda_{\rm s}$CDM is also the Friedmann-Lema\^{i}tre model (see, e.g., Ref.~\cite{Gron07}), but with the exception that the cosmological constant switches sign at certain time $t_\dagger$. For both of the models, the redshift of the matter-radiation equality is given by $1+z_{\rm eq}=2.38\times10^4\Omega_{\rm m0}h^2$. For the $\Lambda$CDM model, $a_{\rm eq}/a_0\sim 3\times 10^{-4}$ (as $z_{\rm eq}\sim3450$ \cite{Planck:2018vyg}), which corresponds to $t_{\rm eq}=\int_{0}^{a_{\rm eq}}(aH)^{-1}\dd{a}\sim5\times 10^4\, \rm yr$. Note that these are negligibly small compared to the present age ($t_0\sim13.8$ Gyr \cite{Planck:2018vyg}) and size ($a_0$) of the Universe, and it is conceivable that this would not change in a viable cosmological model based on $\Lambda_{\rm s}$CDM. Therefore, for our purposes in this section, it will suffice to proceed below by ignoring radiation, namely, by constructing the scale factor of the $\Lambda_{\rm s}$CDM model by gluing (at $t=t_\dagger$) the scale factor of the Friedmann-Lema\^{i}tre model whose cosmological constant is negative (for $t<t_\dagger$), to the one whose cosmological constant is positive (for $t>t_\dagger$). Accordingly, the evolution of the scale factor in the $\Lambda_{\rm s}$CDM model reads
\begin{align}
a(t) =\left\{
  \begin{array}{@{}ll@{}}
    A^{\frac{1}{3}} \sin^{\frac{2}{3}}\qty(\frac{3}{2}\sqrt{\frac{\Lambda_{\rm s0}}{3}}\,t)\quad\quad\quad\quad\text{for}\quad t\leq t_\dagger,\\
   \\
   A^{\frac{1}{3}} \sinh^{\frac{2}{3}}\qty[\frac{3}{2}\qty(\sqrt{\frac{\Lambda_{\rm s0}}{3}}\,t+B)]\,\,\,\text{for}\quad t\geq t_\dagger,
  \end{array}\right. 
  \label{eq:tildezpole}
\end{align}
where
\begin{equation}
\begin{aligned}
A=&\sinh^{-2}\qty[\frac{3}{2}\qty(\sqrt{\frac{\Lambda_{\rm s0}}{3}}\,t_0+B)],
\\
B=&\arcsinh\qty[\sin\qty(\frac{3}{2}\sqrt{\frac{\Lambda_{\rm s0}}{3}}\,t_\dagger)-\frac{3}{2}\sqrt{\frac{\Lambda_{\rm s0}}{3}}\,t_\dagger],
 \end{aligned} \label{eq:where}
\end{equation}
and $t_\dagger<2\pi/\sqrt{3\Lambda_{\rm s0}}$ to ensure $a(t)>0$ for $t>0$. To derive this solution, we have normalized the scale factor such that $a(t_0)=1$ (with $t_0$ being the cosmic time today), and introduced the initial condition $a(0)=0$ (i.e., assumed that the Universe started with a big bang, and used a time parametrization such that the big bang was at $t=0$, which also results in $t_0$ being the age of the Universe today). Note that, under these boundary conditions, general relativity implies, through the Friedmann equations, that this solution satisfies $A=8\pi G \rho_{\rm m0}/{\Lambda_{\rm s0}}$, which also determines the age of the Universe today for a given $\rho_{\rm m0}$ and $\Lambda_{\rm s0}$ using Eq. \eqref{eq:where}. The assumption of an ever-expanding Universe ($H>0$) implies the condition $t_\dagger<\pi/\sqrt{3\Lambda_{\rm s0}}$, as the cosmological constant must switch to its present-day positive value before (in time) the maximum of the sine function is reached. Fig.~\ref{fig:scalefac} illustrates five qualitatively different scenarios varying based on $t_\dagger$. The condition for the ever-expanding Universe, after being used in \eqref{eq:tildezpole} to find the maximum value possible for $a(t_\dagger)=1/(1+z_\dagger)$, translates into the following condition on $z_\dagger$:
\begin{equation}
   z_\dagger> \qty(\frac{\Omega_{\Lambda_{\rm s}0}}{1-\Omega_{\Lambda_{\rm s}0}})^\frac{1}{3}-1.
   \label{eq:positivity1aa}
\end{equation}
Note that \eqref{eq:positivity1aa} can also be easily obtained from \eqref{nlcdm} by enforcing $H>0$ for all redshift values once the radiation density parameter is neglected. If this condition is violated, the Universe enters a contracting phase due to the negative cosmological constant until it switches sign to become positive, which then either restarts the expansion and eventually results in the accelerated expansion of the Universe (dark yellow curve in Fig.~\ref{fig:scalefac}) or further assists the contraction and causes the Universe to recollapse (not present in Fig.~\ref{fig:scalefac}). An effect worth noting for the dark yellow curve in Fig.~\ref{fig:scalefac} is that the one-to-one correspondence between redshift and cosmic time is broken; hence, observations from the same redshift can correspond to signals coming from two different times. We do not elaborate the possibility of these interesting scenarios in the present work. Therefore, in what follows we proceed under the condition of an ever-expanding Universe, which, for instance, gives $z_\dagger>0.33$ for $\Omega_{\Lambda_{\rm s}0}=0.7$.

The deceleration parameter ($q\equiv-\frac{\ddot a}{aH^2}$, where a dot denotes ${\rm d}/{{\rm d}t}$) for the $\Lambda_{\rm s}$CDM model can simply be written as 
\begin{equation}
    q=-1+\frac{3}{2}\left[\frac{\Omega_{\Lambda_{\rm s}0}\,{\rm sgn}[z_\dagger-z]}{1-\Omega_{\Lambda_{\rm s}0}\,{\rm sgn}[z_\dagger-z]}(1+z)^{-3}+1\right]^{-1},
\end{equation}
where we have neglected radiation. For $z>z_\dagger$, it evolves from $q=\frac{1}{2}$ at the matter-dominated epoch toward $q=2$ as the negative cosmological constant dominates with the expansion of the Universe. This equation is solved for $q(z_{\rm c})=0$ only when $z<z_\dagger$, and the solution reads
\begin{equation}
z_{\rm c}=2^{\frac{1}{3}}\left(\frac{\Omega_{\Lambda_{\rm s}0}}{1-\Omega_{\Lambda_{\rm s}0}}\right)^{\frac{1}{3}}-1,
\end{equation}
provided that $z_{\rm c}<z_\dagger$. For the $\Lambda$CDM model, $z_{\rm c}$ is the redshift at which the Universe enters its accelerated phase since its smoothly varying deceleration parameter should pass through the point $q(z_{\rm c})=0$ before becoming negative. For $\Lambda_{\rm s}$CDM, however, due to the discontinuous features of the model, its deceleration parameter does not need to attain the value $q=0$ in order to transit to the accelerated phase from the decelerated phase. While $z_{\rm c}$ defines the redshift at the beginning of the acceleration if $z_{\rm c}<z_\dagger$, if $z_\dagger<z_{\rm c}$, $q=0$ is never satisfied and the deceleration parameter jumps from positive to negative values at $z_\dagger$ which marks the beginning of acceleration in this case (see the dotted gray curve in Fig.~\ref{fig:deceleration} for an example, and see Sec.~\ref{sec:zdagger} for relevant definitions). For example, for $\Omega_{\Lambda_{\rm s}0}=0.7$, in the very extreme case $z_\dagger=0.33$ allowed by Eq.~\eqref{eq:positivity1aa}, $q$ jumps from $\approx 0.82$ to $\approx -0.25$ at $z_\dagger$, and the acceleration begins. Also, the jerk parameter ($j\equiv\frac{\dddot{a}}{aH^3}$) is undefined at the single point $z=z_\dagger$; however, one may check that, when radiation is neglected, both the $\Lambda$CDM and $\Lambda_{\rm s}$CDM models yield $j=1$ everywhere that it is defined throughout the history of the Universe. Note that, if one considers the sign switch feature of $\Lambda_{\rm s}$CDM as an approximation to a DE density that very rapidly yet smoothly transitions from negative to positive, $q$ is not discontinuous and $j$ is not undefined at any point; instead, $q$ goes through a smooth but very sharp transition [e.g., from $q(0.35)\approx 0.8$ to $q(0.33)\approx -0.25$], and $j\gg 1$ during this short transition period while it is again unity (or almost unity) anywhere else.

\subsection{Analyzing the parameter \texorpdfstring{$\bm{z_\dagger}$}{$z_\dagger$}, and its effects on some cosmological tensions}
\label{sec:zdagger}

The deviations of the $\Lambda_{\rm s}$CDM model from the $\Lambda$CDM model are controlled by its additional parameter $z_\dagger$. Before directly confronting the model with observational data in the next section, here we attempt to assess the range and effects of $z_\dagger$. We notice that $\Lambda_{\rm s}$CDM is exactly the same as $\Lambda$CDM at redshifts lower than $z_\dagger$ given that $(\Omega_{\rm m0}h^2)_{\Lambda_{\rm s}\rm CDM}=(\Omega_{\rm m0}h^2)_{\Lambda\rm CDM}$ and $\Lambda_{\rm s0}=\Lambda$, while these two models differ at redshifts larger than $z_\dagger$ as $\Lambda_{\rm s}(z>z_\dagger)=-\Lambda_{\rm s0}$ in $\Lambda_{\rm s}$CDM, yet this difference disappears once again at even larger redshifts, as the corresponding density parameters, $\Omega_{\Lambda_{\rm s}}=\Lambda_{\rm s}/(3 H^2)$ and $\Omega_{\Lambda}=\Lambda/(3 H^2)$, regardless of whether they yield positive or negative values, rapidly become negligible with increasing redshift in both models. Thus, $\Lambda_{\rm s}$CDM differs from $\Lambda$CDM for $z_\dagger<z\ll z_*$; hence, it is, in practice, a post-recombination modification to $\Lambda$CDM. However, note that the abrupt-change feature of $H(z)$ in $\Lambda_{\rm s}$CDM (or of the models that are well approximated by such as the gDE) would not be captured by the spline reconstruction of the Hubble parameter in Refs.~\cite{Bernal:2016gxb,Aylor:2018drw,Knox:2019rjx}; hence, it evades their arguments against post-recombination deviations from $\Lambda$CDM, and furthermore, since $j(z)=1$ (neglecting radiation) and we expect $q_0\sim-0.55$ at $z\sim 0$ for $\Lambda_{\rm s}$CDM as in $\Lambda$CDM, a direct comparison of its $H_0$ value with the SH0ES Collaboration measurements of $H_0$ \cite{Riess:2016jrr,Riess:2019cxk}
should not be an issue, unlike models with rapidly changing $H(z)$ values for $z\lesssim0.1$ \cite{Efstathiou:2021ocp,Camarena:2021jlr}. The SH0ES $H_0$ determination is a two-step process: first, anchors, Cepheids, and calibrators are combined to produce a constraint on the type Ia supernovae (SnIa) absolute magnitude $M_B$, and second, Hubble-flow SnIa data are used to probe the luminosity distance-redshift relation in order to determine $H_0$ by adopting a cosmography with $q_0=-0.55$ and $j_0=1$ \cite{Riess:2016jrr} (small deviations from $q_0=-0.55$ have an insignificant effect on the determined $H_0$ value \cite{Riess:2020fzl,Camarena:2021jlr}). These suggest that, as $\Lambda_{\rm s}$CDM yields $q_0\sim-0.55$ (see Fig.~\ref{fig:deceleration}) and $j_0=1$, it respects the methodology used by the SH0ES Collaboration to obtain $M_B$ and $H_0$; thus, if $\Lambda_{\rm s}$CDM is to resolve the SH0ES $H_0$ tension, it is conceivable that it will also be in good agreement with the SH0ES $M_B$ measurement \cite{Camarena:2021jlr,Camarena:2019moy}.

We now analyze the parameter $z_\dagger$ with respect to the $H_0$, Ly-$\alpha$ and Galaxy BAO measurements while the consistency with the CMB data is ensured. To do so, we fix the comoving angular diameter distance to last scattering, $D_M(z_{*})$, to that of $\Lambda$CDM for $\Lambda_{\rm s}$CDM (we assume $z_{*}=1100$ for both models). This is a good guiding principle since once the sound horizon at CMB last scattering, $r_{*}$, is given, $D_M(z_{*})$ is very strictly constrained in an almost model-independent way by the measurement of the angular acoustic scale $\theta_{*}$ since $D_M(z_{*})=r_{*}/\theta_*$. And, for  $\Lambda_{\rm s}$CDM, we expect almost no deviations in the pre-recombination dynamics of the Universe, and hence in $r_{*}$, once we fix its $\rho_{\rm m}(z_{*})$ and $\rho_{\rm r}(z_*)$ values to those of $\Lambda$CDM. Fixing $\rho_{\rm m}(z_{*})$ in this way is well justified as this value is very well constrained by the relative heights of the CMB power spectra peaks, and its corresponding baryon density is in good agreement with standard big bang nucleosynthesis (BBN), providing even more confidence. Since $\rho_{\rm r}(z_*)$ is also fixed by the CMB monopole temperature measurements, the only difference regarding the pre-recombination dynamics would be due to the difference between the values of $\Lambda_{\rm s}$ in $\Lambda_{\rm s}$CDM and $\Lambda$ in $\Lambda$CDM, but, since these have negligible corresponding energy densities for $z\geq z_{*}$, $r_{*}$ is not significantly affected. We fix $z_{*}=1100$ simply because it is a reasonable choice and we do not expect it to affect our argumentation since the relevant integrals are not substantially affected by its sensible deviations. After we fix $D_M(z_{*})$ in this way, we can calculate $\Lambda_{\rm s0}$ using the equality $D_{M}(z)=c\int_0^{z}H^{-1}(z')\dd{z'}$ for the comoving angular diameter distance at $z$, which is satisfied for the spatially flat Robertson-Walker (RW) metric. Knowing  $\Lambda_{\rm s0}$, $\rho_{\rm m}$ and $\rho_{\rm r}$ at a single point allows us to construct $H(z)$ at all times and discuss how $z_\dagger$ modifies $H(z)$ and $H_0$ with respect to observations using visualization methods similar to those of Ref.~\cite{Aubourg:2014yra}.
 
 \begin{figure}[t!]
  \includegraphics[width=0.48\textwidth]{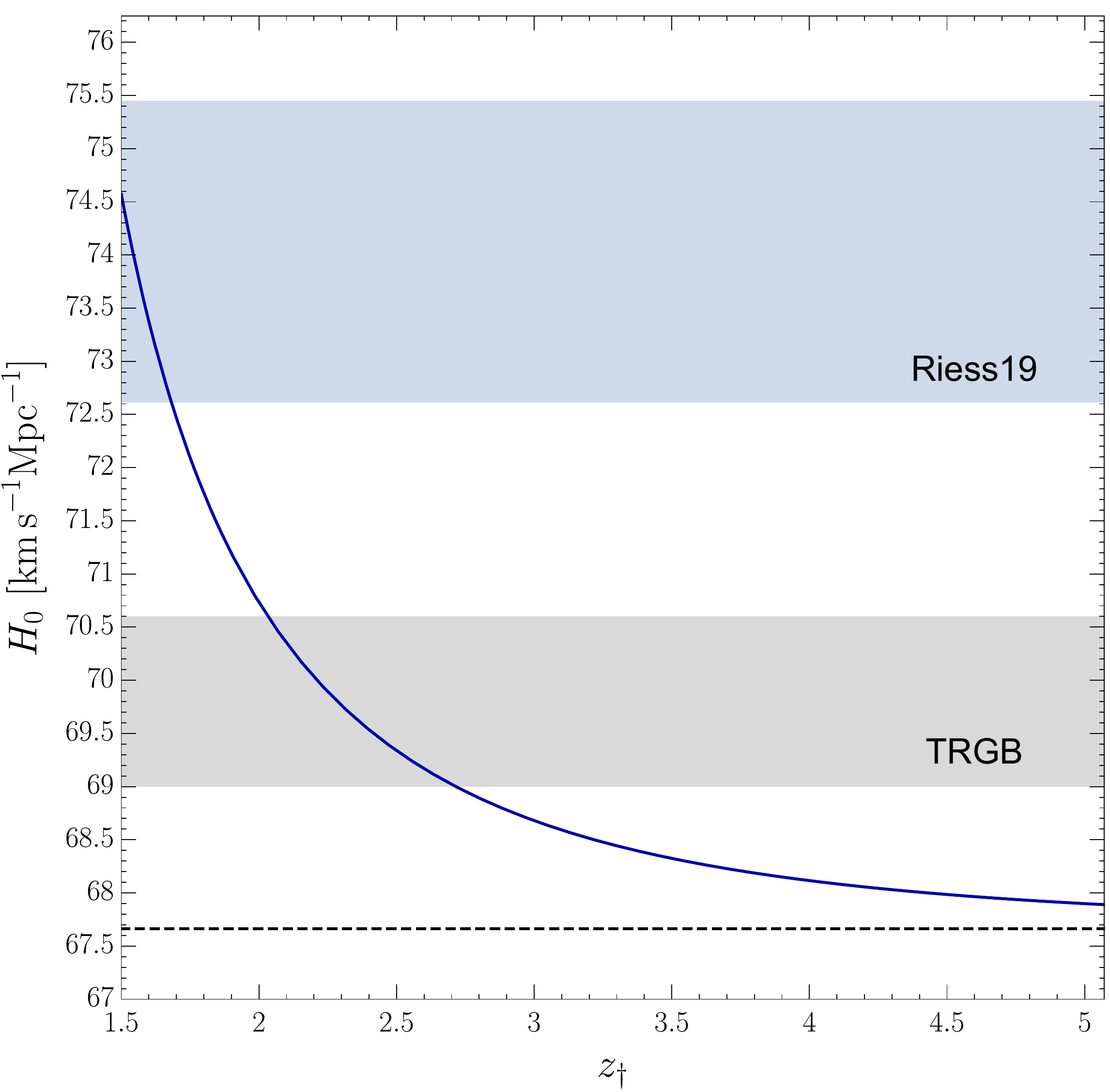}
  \caption{$H_0$ versus $z_\dagger$ for the $\Lambda_{\rm s}$CDM model (solid curve), and the $\Lambda$CDM model (dashed line). The values are calculated by fixing $D_M(z_{*})$ and $\rho_{\rm m}(z_{*})$ (and hence $\rho_{\rm m0}$) to that of $\Lambda$CDM using the mean values of the \textit{Planck} 2018 TT,TE,EE+lowE+lensing results \cite{Planck:2018vyg}. The gray band is the model-independent TRGB measurement $H_0= 69.8 \pm 0.8\, {\rm km\,s}^{-1}{\rm Mpc}^{-1}$ \cite{Freedman:2019jwv} and the blue band is the Cepheid measurement $H_0=74.03\pm1.42\, {\rm km\,s}^{-1}{\rm Mpc}^{-1}$ \cite{Riess:2019cxk}.
  }
  \label{fig:zstar_vs_H0}
\end{figure}

 \begin{figure}[t!]
  \includegraphics[width=0.48\textwidth]{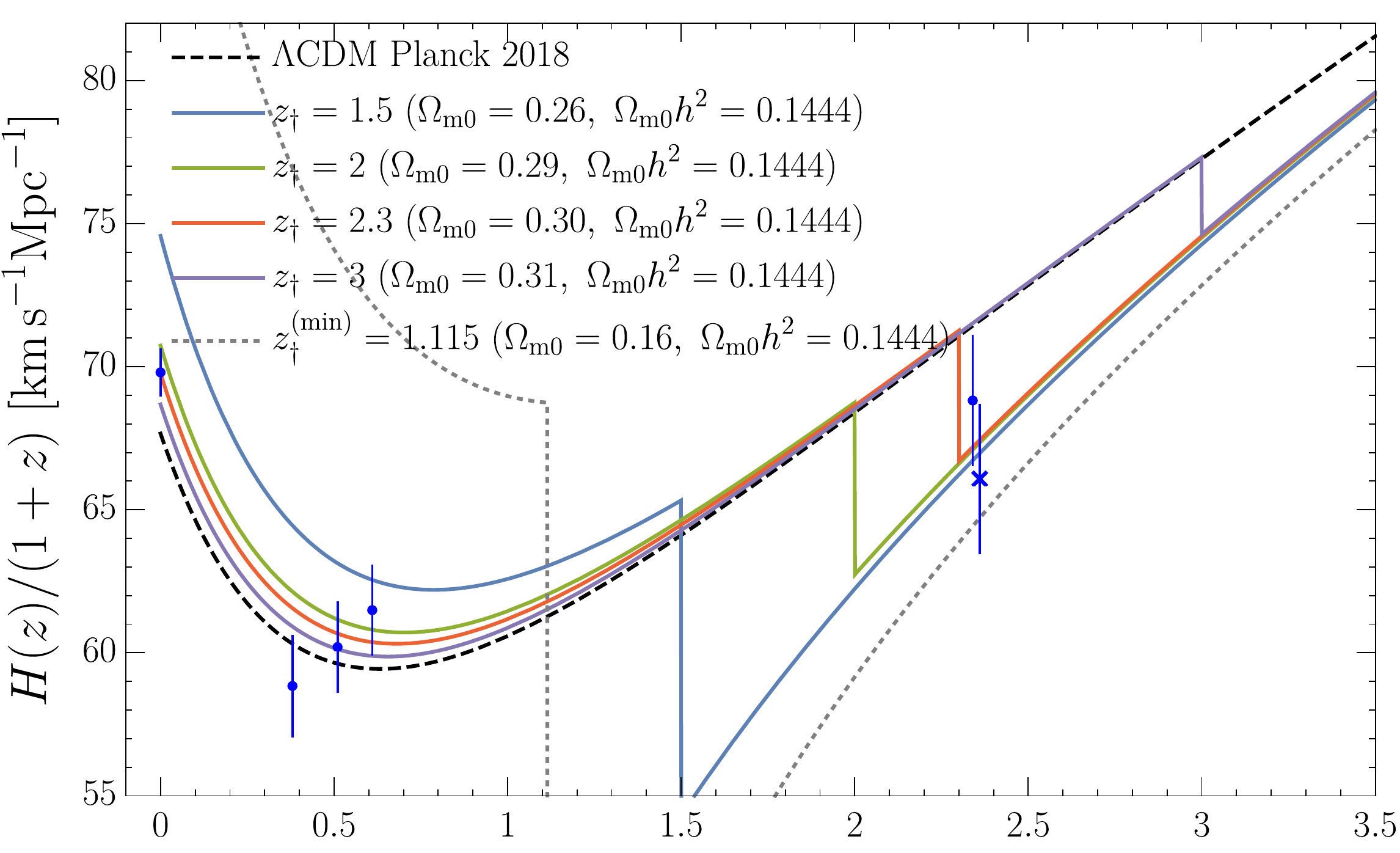}
  \includegraphics[width=0.48\textwidth]{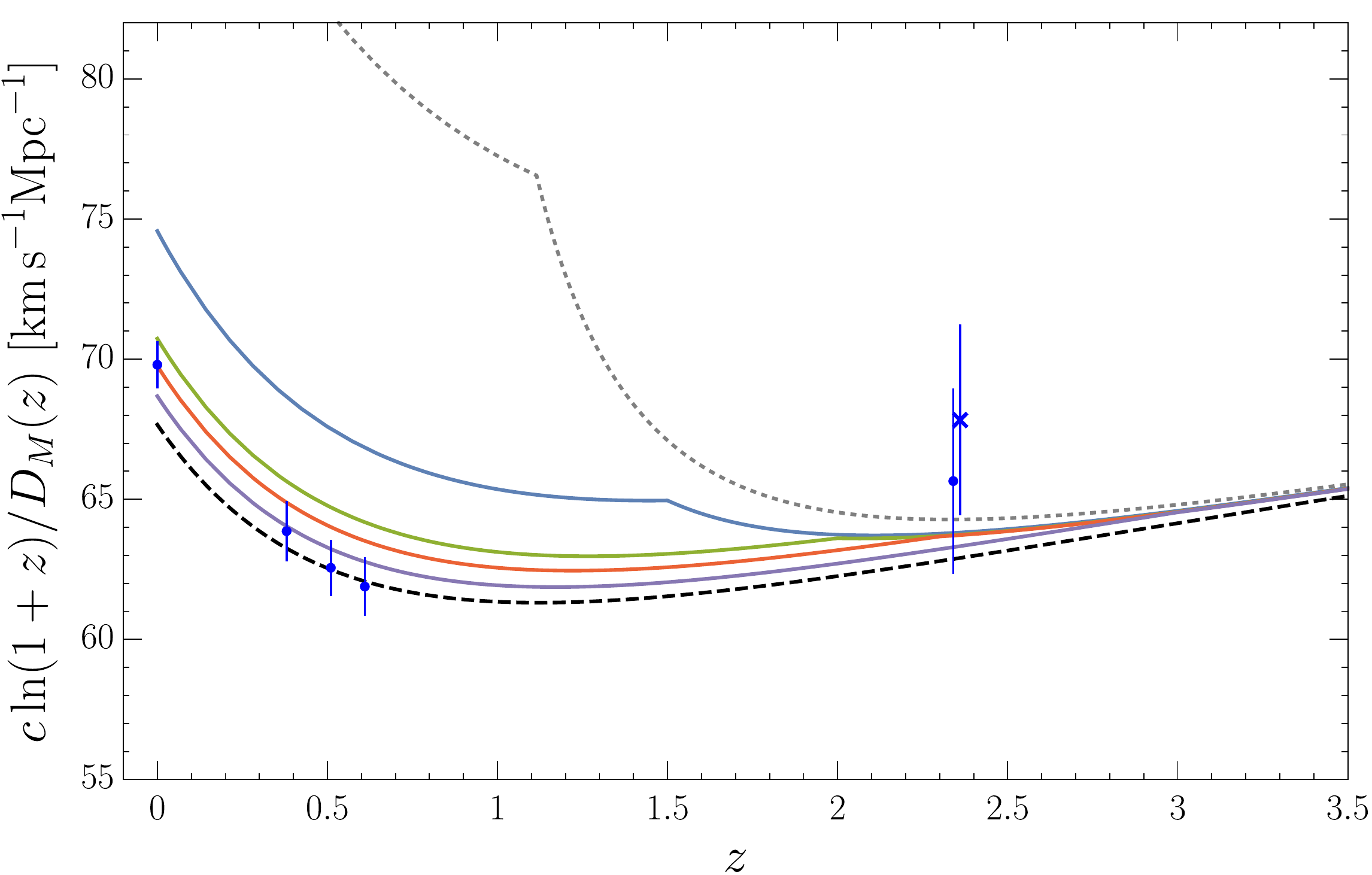}
  \caption{Comoving Hubble parameter and the comoving angular diameter distance versus redshift for various $z_\dagger$ values for the $\Lambda_{\rm s}$CDM model. All of the plots are drawn by fixing $D_M(z_*)$ and $\rho_{\rm m}(z_*)$ (and hence fixing $\rho_{\rm m0}$) to that of the $\Lambda$CDM model using mean values of the \textit{Planck} 2018 TT,TE,EE+lowE+lensing results. We consider the observational $H(z)$ values (blue error bars), $H_0=69.8\pm0.8\,{\rm km\,s}^{-1}{\rm Mpc}^{-1}$ from the TRGB \cite{Freedman:2019jwv}, consensus Galaxy BAO (from $z_{\rm eff}=0.38,\,0.51,\,0.61$) and DR14 Ly-$\alpha$ BAO (from $z_{\rm eff}=2.34,\,2.35$) \cite{Blomqvist:2019rah,Agathe:2019vsu,Anderson:2013zyy}. }
  \label{fig:prelim_evo}
\end{figure}

This construction is done in Figs.~\ref{fig:zstar_vs_H0} and \ref{fig:prelim_evo} based on the results of \textit{Planck} 2018 \cite{Planck:2018vyg} (see the figure captions for more details) but neglecting the radiation energy density. It is seen from Fig.~\ref{fig:zstar_vs_H0} that $\Lambda_{\rm s}$CDM attains greater values of $H_0$ compared to $\Lambda$CDM, and $z_\dagger$ is inversely correlated with $H_0$. Such greater values are a direct consequence of the sudden drop in $H(z)$ due to the negative cosmological constant for $z>z_\dagger$ as explained in the Introduction. Additionally, as seen in the top panel of Fig.~\ref{fig:prelim_evo}, the drop in $H(z)$ due to the sign switch allows $\Lambda_{\rm s}$CDM to better agree with the Ly-$\alpha$ data; however, this amelioration of the Ly-$\alpha$ discrepancy disappears immediately for $z_\dagger\gtrsim 2.4$. Moreover, as $z_\dagger$ increases, $H_0$ decreases, approaching the value of $\Lambda$CDM as $z_\dagger\to\infty$. This is because of two reasons: first, as $z_\dagger$ increases, the portion of the $D_M(z_{*})$ integral that is over negative values of $\Lambda_{\rm s}$ decreases and hence requires less compensation from the positive $\Lambda_{\rm s}$ portion including $H_0$; second, as $z_\dagger$ increases, the sign-switching feature of $\Lambda_{\rm s}$ becomes rapidly less effective since, for large $z_\dagger$, matter is the dominant energy component of the Universe at the time of the sign switch and the effect of negative $\Lambda_{\rm s}$ on the evolution of $H(z)$ is negligible. If we consider $z_\dagger=3$, just before the cosmological constant becomes negative ($z\to z_\dagger^-$), the matter already is by far the dominant component of the Universe, viz., $\Omega_{\rm m}(z=3)\approx0.96$ corresponding to only $\abs{\Omega_{\Lambda_{\rm s}}/\Omega_{\rm m}}\approx0.04$. It is intriguing that, for $z_\dagger=2.3$, which is almost as high as $z_\dagger$ can get without losing the improved agreement with the Ly-$\alpha$ data, the $H_0$ value is in excellent agreement with $H_0= 69.8 \pm 0.8\, {\rm km\,s}^{-1}{\rm Mpc}^{-1}$~\cite{Freedman:2019jwv} (revised as $H_0= 69.6 \pm 0.8\, {\rm km\,s}^{-1}{\rm Mpc}^{-1}$ in Ref.~\cite{Freedman:2020dne}) from a recent calibration of the tip of the red giant branch (TRGB) applied to type Ia supernovae. Both of these effects on $H_0$ and $H(z\approx 2.34)$ suggest that $\Lambda_{\rm s}$CDM might be most effective for $z_\dagger\lesssim 2.34$. In line with this, as Fig.~\ref{fig:zstar_vs_H0} demonstrates, $H_0$ is greater for smaller values of $z_\dagger$; for $z_\dagger=1.5$, $H_0$ goes up to $\approx74.5~{\rm km\, s^{-1}\, Mpc^{-1}}$, so $z_\dagger>1.5$ covers all the recent local measurements of $H_0$, including the largest $H_0$ estimations by the SH0ES Collaboration (see Refs.~\cite{Riess:2016jrr,Riess:2018byc,Riess:2019cxk,Freedman:2019jwv,Yuan:2019npk,Freedman:2020dne,Riess:2020fzl}). However, looking at the bottom panel of Fig.~\ref{fig:prelim_evo}, we see that as $z_\dagger$ gets smaller, a greater tension with the comoving angular diameter distance measurements from Galaxy BAO data is generated. In fact, Fig.~\ref{fig:prelim_evo} seems to suggest that the smaller the value of $z_\dagger$, the greater the tension with the Galaxy BAO data, and the extent of this effect in limiting the increase in $H_0$ is not clear without a robust observational analysis.

\begin{figure}[t!]
    \centering
    \includegraphics[width=0.48\textwidth]{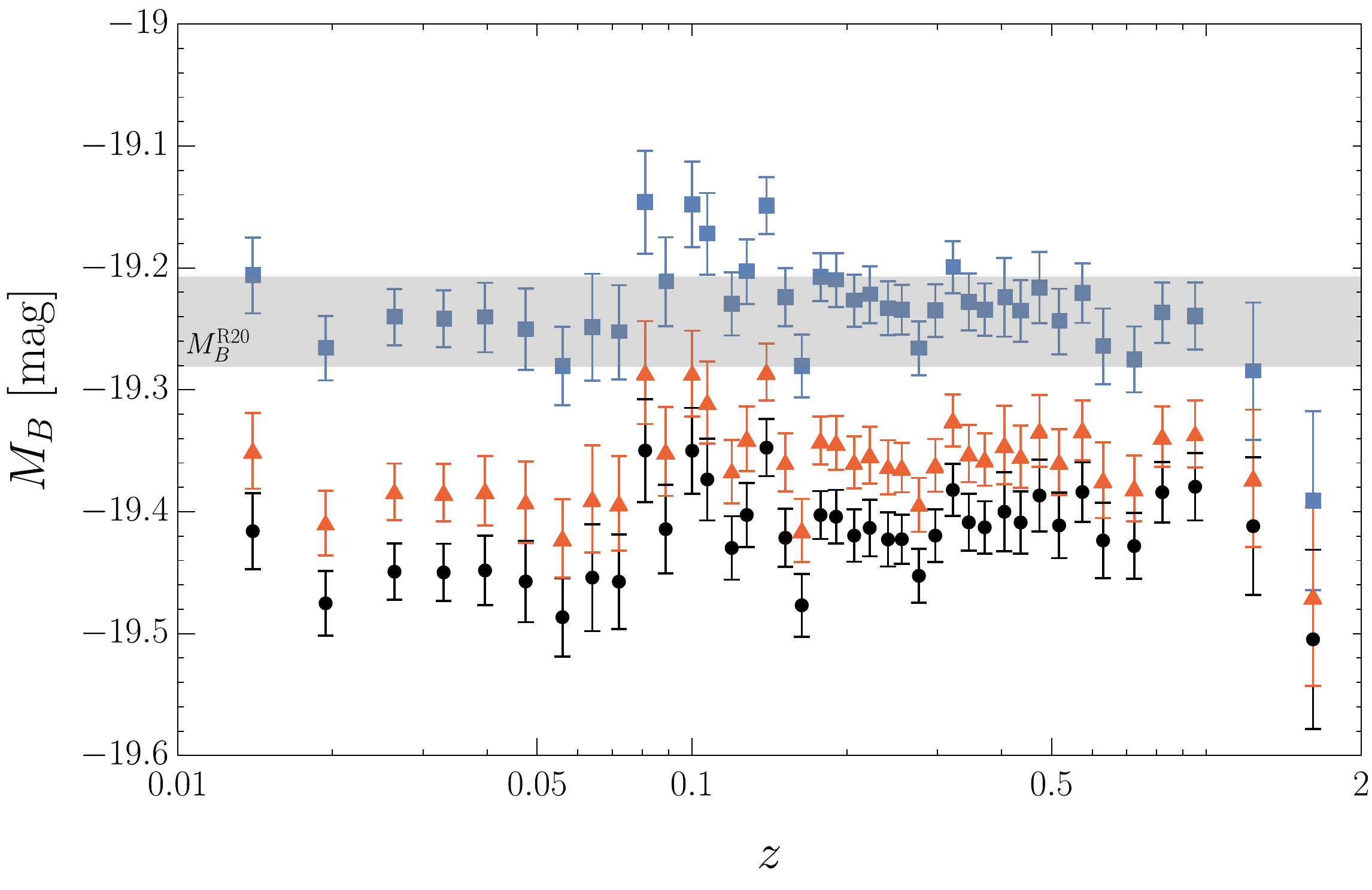}
    \caption{Inferred SnIa absolute magnitudes $M_{B,i}=m_{B,i}-\mu(z_i)$ of the binned Pantheon sample containing SnIa apparent magnitudes $m_{B,i}$ (with 68$\%$ C.L. error bars) \cite{Scolnic:2017caz} for the distance moduli $\mu(z_i)$ assuming $z_\dagger=1.5$ (blue) (which is in excellent agreement with the SH0ES $H_0$ value), $z_\dagger=2.3$ (red) (which is in excellent agreement with the TRGB $H_0$ value), and $\Lambda$CDM \textit{Planck} 2018 (black), all calculated using the corresponding $H(z)$ functions given in Fig.~\ref{fig:prelim_evo} with matching colors. The grey bar is the 68$\%$ C.L. constraint from Cepheid calibrations \cite{Camarena:2021jlr}.}
    \label{fig:prelimMB}
\end{figure}

The discrepancy of the latest SH0ES $H_0$ determination $H_0^{\rm R20}=73.2\pm1.3\, {\rm km\,s}^{-1}{\rm Mpc}^{-1}$ \cite{Riess:2020fzl} and $\Lambda$CDM \textit{Planck} 2018 constraint $H_0=67.36\pm0.54\, {\rm km\,s}^{-1}{\rm Mpc}^{-1}$ \cite{Planck:2018vyg} is equivalent to the discrepancy of the Pantheon SnIa absolute magnitudes, which have a value $M^{\rm Planck}_B=-19.401\pm0.027\,{\rm mag}$ 
\cite{Camarena:2019rmj} when calibrated using the CMB sound horizon and propagated via BAO measurements to low $z$ (inverse distance ladder, $z\simeq 1100$), in significant tension ($3.4\sigma$) with the value $M_B^{\rm R20}=-19.244\pm0.037\,{\rm mag}$ \cite{Camarena:2021jlr} (using Pantheon SnIa data set \cite{Scolnic:2017caz}) when the calibration is done using Cepheid stars at $z<0.01$. This tension is reflected in the inferred SnIa absolute magnitudes from $M_{B,i}=m_{B,i}-\mu(z_i)$ [where $\mu(z_i)=5\log_{10}\big[\frac{1+z_i}{10\,{\rm pc}}\int_0^{z_i}\frac{c\dd{z}}{H(z)}\big]$ is the distance modulus for the spatially flat RW metric and $m_{B,i}$ is the measured apparent magnitude of the supernovae at redshift $z_i$ ($z_i>0.01$)] using the distance modulus corresponding to the $\Lambda$CDM \textit{Planck} 2018 curve in Fig.~\ref{fig:prelim_evo}, which are in tension with $M_B^{\rm R20}$ from Cepheid calibrators (see black error bars in Fig.~\ref{fig:prelimMB} and the caption of the figure for information about the $m_{B,i}$ data that we used). On the other hand, we see from the figure that, for $z_\dagger=2.3$ (red bars) (i.e., when $\Lambda_{\rm s}$CDM agrees with the TRGB $H_0$ measurement) the inferred $M_{B,i}$ values are systematically shifted upwards, relaxing the tension with $M_B^{\rm R20}$, and for $z_\dagger=1.5$ (blue bars) (i.e., when $\Lambda_{\rm s}$CDM agrees with the SH0ES $H_0$ measurement) the estimated absolute magnitudes from $\Lambda_{\rm s}$CDM are in excellent agreement with $M_B^{\rm R20}$. It is no surprise that $\Lambda_{\rm s}$CDM results in greater $M_{B,i}$ values compared to $\Lambda$CDM for $z<z_\dagger$, because it is guaranteed that, compared to $\Lambda$CDM with the same $D_M(z_*)$ and $\Omega_{m0}h^2$ values, $\Lambda_{\rm s}$CDM has greater $H(z<z_\dagger)$ values making its $\mu(z<z_\dagger)$ smaller. A subtler point is that, although $H(z>z_\dagger)$ is smaller for $\Lambda_{\rm s}$CDM, it will keep resulting in greater $M_{B,i}$ values up to $z\sim z_*$ since the smaller value of the $\mu(z)$ of $\Lambda_{\rm s}$CDM catches up to that of $\Lambda$CDM only at the redshift to which their angular diameter distance is equal, i.e., at last scattering for which $D_M(z_*)$ is the same among these models. In addition, since smaller $z_\dagger$ values amplify the above-mentioned deviance of $\Lambda_{\rm s}$CDM, $M_{B,i}$ are inversely correlated with $z_\dagger$ just as $H_0$ is. An important point is that $\Lambda_{\rm s}$CDM not only systematically results in higher $M_{B,i}$ values, but also respects the internal consistency of the SH0ES measurements by simultaneously matching their $H_0$ and $M_{B}$ constraints \cite{Riess:2016jrr,Riess:2018byc,Riess:2019cxk,Riess:2020fzl,Camarena:2021jlr,Camarena:2019moy}. This is not true in general for models with deviations from $\Lambda$CDM at low redshifts, e.g., models with a dynamical DE equation-of-state parameter, or models of smoothly nonminimally interacting DE \cite{Efstathiou:2021ocp,Camarena:2021jlr,Benevento:2020fev,Lemos:2018smw,Alestas:2021xes,DeFelice:2020cpt}; however, see Ref.~\cite{Nunes:2021zzi} for an analysis in this context excluding CMB data, and Refs.~\cite{Theodoropoulos:2021hkk,Alestas:2020zol,Marra:2021fvf,Perivolaropoulos:2021bds} for astrophysical (rather than cosmological) approaches addressing the $M_B$ tension.

\begin{figure}[b!]
    \centering
    \includegraphics[width=0.48\textwidth]{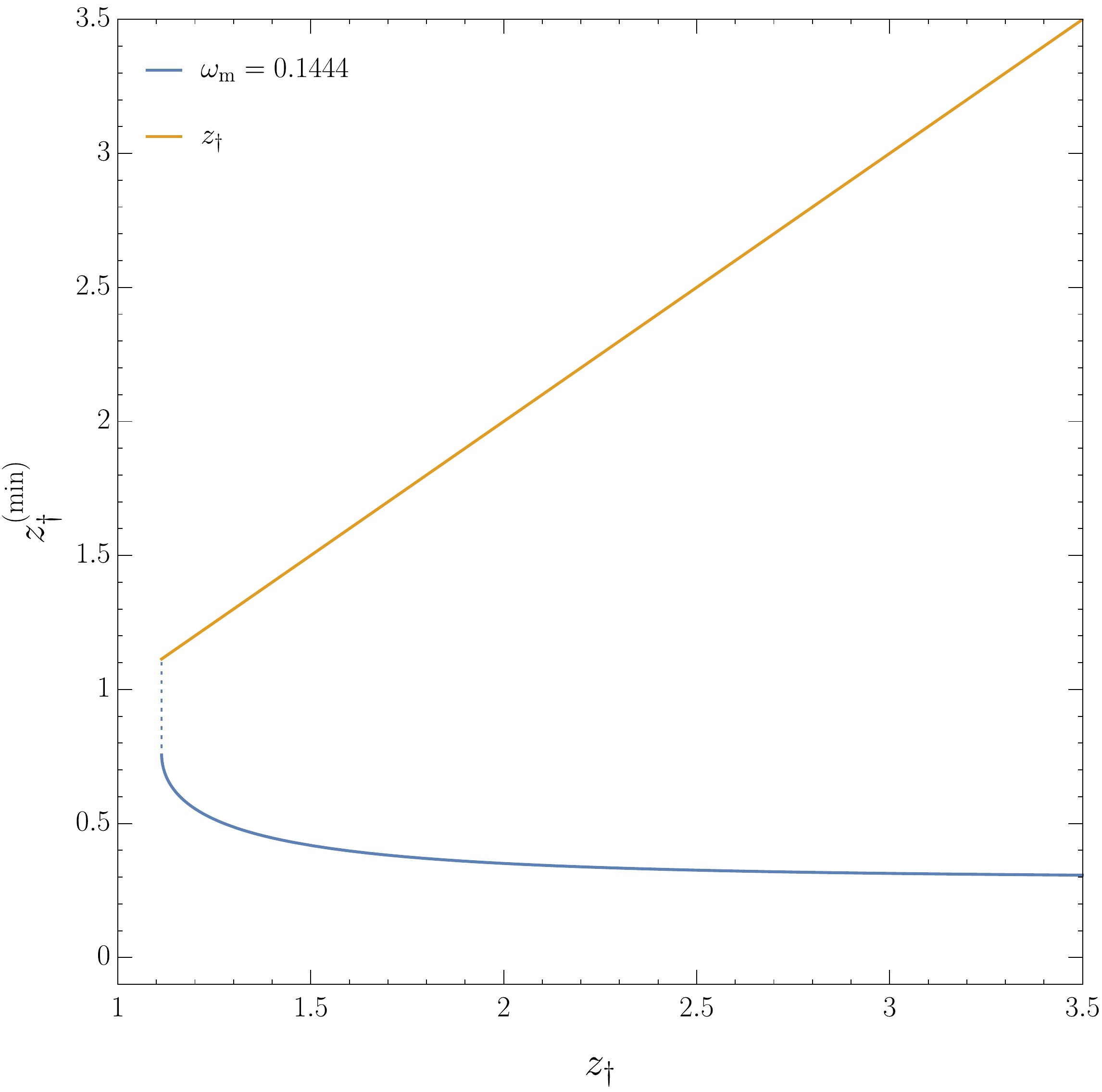}
    \caption{We solve numerically that $z_\dagger^{\rm (min)}\approx1.1$. The point of intersection of the straigt line (orange) and the curve (blue), is the solution of Eq.~\eqref{eq:positivity1a}.}
    \label{fig:zmin}
\end{figure}

\begin{figure}[t!]
    \centering
    \includegraphics[width=0.48\textwidth]{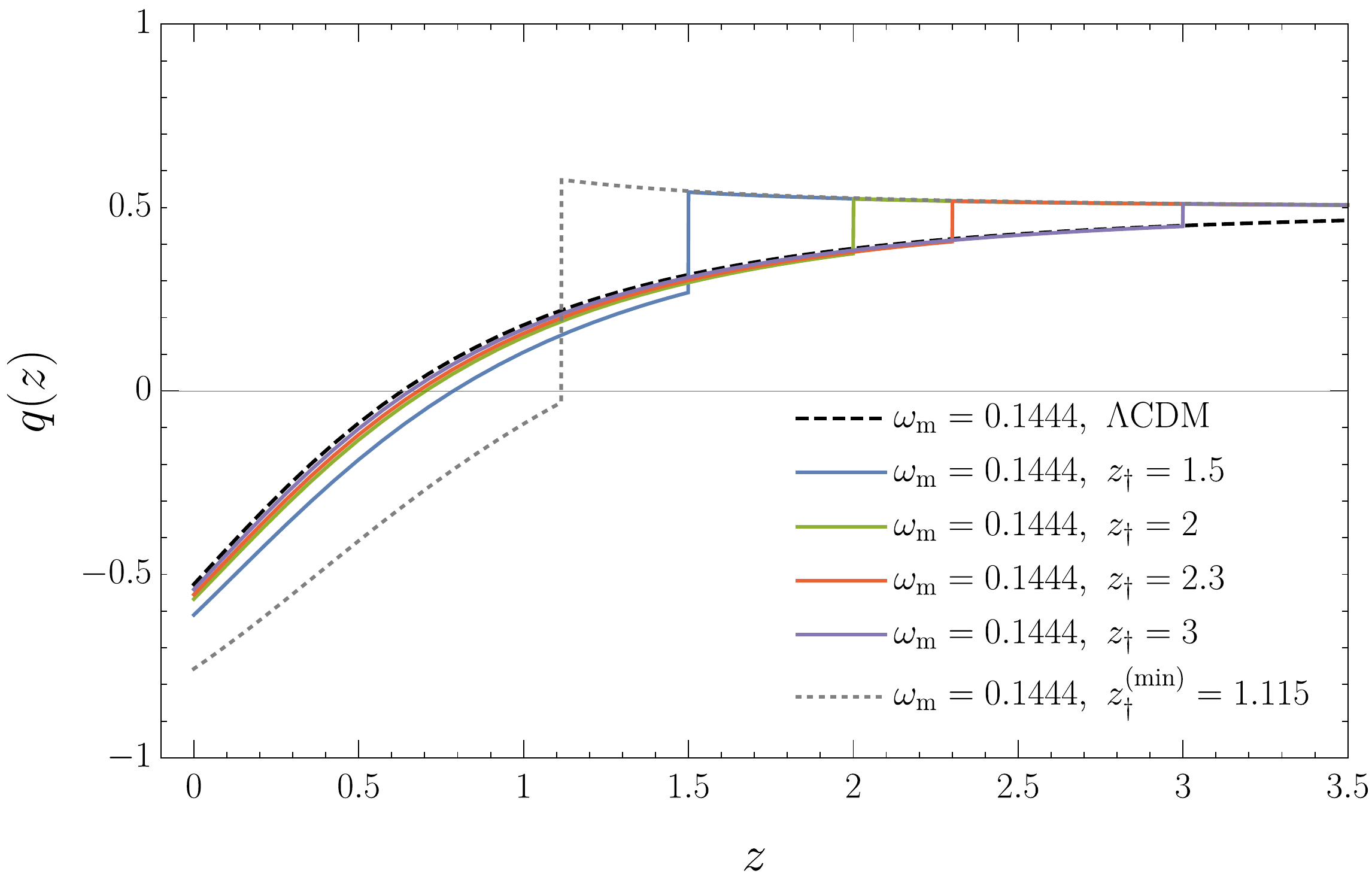}
    \caption{Evolution of the deceleration parameter $q(z)$ for various $z_\dagger$ values, including $z_\dagger\approx z_{\dagger}^{\rm (min)}$, corresponding to Fig.~\ref{fig:prelim_evo}.
    }
    \label{fig:deceleration}
\end{figure}

As a final remark for this section, we notice that the condition for an ever-expanding Universe given in Eq. \eqref{eq:positivity1aa} implies
\begin{equation}
   z_\dagger^{\rm (min)}= \qty(\frac{h_{\rm (max)}^2}{\omega_{\rm m}}-1)^\frac{1}{3}-1,
   \label{eq:positivity1a}
\end{equation}
where $\omega_{\rm m}\equiv\Omega_{\rm m0}h^2\propto\rho_{\rm m0}$ and $h^2_{(\rm max)}$ is the maximum $h$ value attainable  while satisfying the constraint on $D_M(z_*)$ by the ever-expanding $\Lambda_{\rm s}$CDM Universe for a given $\omega_{\rm m}$. This also determines $\Omega_{\rm m}^{\rm (min)}$, and thereby $\Omega_{\Lambda}^{\rm (max)}$ as well. We solve numerically that $z_\dagger^{\rm (min)}\approx1.1$ for $\omega_{\rm m}=0.1444$ (this value is chosen based on \textit{Planck} 2018 \cite{Planck:2018vyg} as in Fig.~\ref{fig:prelim_evo}); see Fig.~\ref{fig:zmin}. We plot the deceleration parameter in Fig.~\ref{fig:deceleration} for $z_\dagger$ values, including $z_\dagger\approx z_\dagger^{\rm (min)}$ for which the acceleration starts at $z_\dagger$ and not $z_{\rm c}$. It is astonishing that even for this extreme value $z_\dagger=1.115$, which is approximately the limit of the ever-expanding Universe condition we obtained while ensuring the consistency with the \textit{Planck} CMB data, the good representation of the Ly-$\alpha$ data remains, as seen in Fig.~\ref{fig:prelim_evo}. This shows that it is an intrinsic feature of the $\Lambda_{\rm s}$CDM scenario, which provides an AdS background for $z>z_\dagger$, to be consistent with the available cosmological data from $z\gtrsim1$.

To summarize, the $\Lambda_{\rm s}$CDM model has the potential to resolve both the $H_0$ and $M_B$ tensions while remaining consistent with the CMB data; the pre-recombination physics were practically untouched in this analysis. The model comes with the additional benefit of better agreeing with the Ly-$\alpha$ measurements for $z_\dagger \lesssim 2.34$. However, the comoving angular diameter distance measurements from Galaxy BAO oppose the amelioration in $H_0$ and $M_{B,i}$ by insisting that $z_\dagger$ does not attain very small values. This opposition may permit a partial alleviation of the $H_0$ tension rather than its resolution when, e.g., $H_0=74.03\pm1.42\, {\rm km\,s}^{-1}{\rm Mpc}^{-1}$ from the Cepheid measurement of $H_0$ \cite{Riess:2019cxk} is considered; however, it may allow for a full resolution if one considers $H_0= 69.8 \pm 0.8\, {\rm km\,s}^{-1}{\rm Mpc}^{-1}$ from the TRGB measurement of $H_0$ \cite{Freedman:2019jwv}, which might prove to be sufficient with forthcoming observations. There appears to be an interval $1.5\lesssim z_\dagger\lesssim2.34$ where the comoving angular diameter distance data of Galaxy BAO can reconcile with the Ly-$\alpha$ BAO and $H_0$ measurements within $\Lambda_{\rm s}$CDM. The observational analysis in the next section will reveal how efficient the features of the $\Lambda_{\rm s}$CDM model can work to alleviate the tensions prevailing in the standard cosmological model when confronted with data.

\section{Observational constraints and results}
 \label{sec:obs}

\begin{table*}[ht!]
  \caption{Constraints (68\% C.L.) on  the free and some derived parameters of the $\Lambda_{\rm s}$CDM and standard $\Lambda$CDM models for CMB and CMB+BAO data. The parameter $H_{\rm 0}$ is measured in units of km s${}^{-1}$ Mpc${}^{-1}$. In the last three rows, the best fit ($-2\ln{\mathcal{L}_{\rm max}}$), the $\log$-Bayesian evidence ($\ln \mathcal{Z}$), and the relative $\log$-Bayesian evidence $\Delta\ln \mathcal{Z}=\ln \mathcal{Z}_{\rm reference}-\ln \mathcal{Z}$ are listed.}
  \label{tab:priors}
	\scalebox{0.94}{
	\setlength\extrarowheight{2pt}
	\begin{centering}
	  \begin{tabular}{lcccccc}
  	\hline
    \toprule
    \multicolumn{1}{l}{\textbf{Data set}} & \multicolumn{3}{c}{\textbf{CMB}} & \multicolumn{3}{c}{\textbf{CMB+BAO}} \\  \hline
      & \textbf{{$\bm{\Lambda}$CDM}} & \textbf{$\bm{\Lambda}_{\textbf{s}}$CDM} & \textbf{$\bm{\Lambda}_{\textbf{s}}$CDM$\bm{+z_\dagger=2.32}$} & \textbf{{$\bm{\Lambda}$CDM}} & \textbf{$\bm{\Lambda}_{\textbf{s}}$CDM} & \textbf{$\bm{\Lambda}_{\textbf{s}}$CDM$\bm{+z_\dagger=2.32}$}  \\ 
      \midrule
      \vspace{0.1cm}
$10^{2}\omega_{\rm b }$ & $2.235\pm0.015$ &  $2.238 \pm 0.015$ & $2.238\pm 0.015$ &   $2.244 \pm 0.013$ 
    & $2.231 \pm 0.014$    & $2.230\pm 0.013$  \\ 
    
\vspace{0.1cm}
$\omega_{\rm c }$  &$0.1201 \pm 0.0014$ & $0.1197 \pm 0.0013$ & $0.1199\pm0.0013$ & $0.1189\pm 0.0009$ 
 & $0.1208 \pm 0.0011$ & $0.1209 \pm 0.0009$  \\ 
 
\vspace{0.1cm}
$100 \theta_{s } $  & $1.04090 \pm 0.00031$    &$1.04093\pm 0.00030$  & $1.04091\pm 0.00031$ & $1.04102\pm 0.00029$   
&    $1.04081 \pm 0.00029$& $1.04080 \pm 0.00029$  \\ 

\vspace{0.1cm}
$\ln(10^{10}A_{\rm s})$ &     $3.044\pm0.016$  & $3.043\pm0.016$ & $3.043\pm 0.016$ & $3.045\pm0.016$ & $3.043\pm0.016$  &  $3.043\pm 0.016$  \\ 

\vspace{0.1cm}
$n_{s } $  & $0.9646\pm0.0043$   & $0.9657\pm 0.0044$  & $0.9655\pm 0.0044$ & $0.9673\pm 0.0037$  
&     $0.9633\pm 0.0039$ & $0.9632 \pm 0.0036$  \\ 

\vspace{0.1cm}
$\tau_{\rm reio } $  &     $0.0543\pm0.0078$  & $0.0542\pm0.0078$ & $0.0541 \pm 0.0078$ & $0.0559\pm 0.0078$ 
& $0.0530\pm 0.0077$  &  $0.0526 \pm 0.0075$   \\ 

\vspace{0.1cm}
$z_\dagger$  &   --- & unconstrained  & $[2.32]$ & --- 
& $2.44\pm 0.29$ & $[2.32]$    \\
\hline

\vspace{0.10cm} 
$\Omega_{\rm{m} }$ &   $0.3162 \pm 0.0084$   & $0.2900 \pm 0.0160$  & $0.2967 \pm 0.0086$ & $0.3090\pm 0.0059$ & $0.3035\pm 0.0062$  & $0.3029 \pm 0.0060$ \\

\vspace{0.10cm} 
$H_{\rm 0}$ &    $67.29\pm0.60$  & $70.22\pm1.78$   & $ 69.42 \pm 0.71$ & $67.81\pm0.44$  & $68.82\pm0.55$   & $68.91 \pm 0.48$  \\

\vspace{0.10cm} 
$\sigma_{8}$ &    $0.8117\pm 0.0076$  & $0.8223 \pm 0.0098$   & $0.8186 \pm 0.0074$ & $0.8090 \pm 0.0073$  & $0.8207 \pm 0.0080$   & $0.8215 \pm 0.0071$  \\

  \vspace{0.10cm} 
$S_{8}$ &    $0.8332\pm 0.0163$  & $0.8071\pm 0.0210$   & $0.8138 \pm 0.0166$ & 
$0.8219 \pm 0.0127$  & $0.8255 \pm 0.0128$   & $0.8264 \pm 0.0126$  \\
  \hline
 
 \vspace{0.10cm} 
   $-2\ln{\mathcal{L}_{\rm max}}$ & $1386.52$ & $1385.73$ & $1386.56$ & $1394.32$ & $1393.77$ & $1393.54$\\ 

\vspace{0.10cm} 
   $\ln \mathcal{Z}$ & $-1424.19$ & $-1424.22$ & $-1423.50$ & $-1431.46$ & $-1432.77$ & $-1431.89$\\ 

\vspace{0.10cm} 
   $\Delta\ln \mathcal{Z}$ & $0.69$ & $0.72$ & $0$ & $0$ & $1.31$ & $0.43$\\ 

    \bottomrule
    \hline 
  \end{tabular}
  \end{centering}
  }
\end{table*}

Considering the background and perturbation dynamics, in what follows we explore the full parameter space of the $\Lambda_{\rm s}$CDM model and, for comparison, that of the standard $\Lambda$CDM model. The baseline seven free parameters of the $\Lambda_{\rm s}$CDM model are:
\begin{equation}
\label{baseline1}
\mathcal{P}= \left\{ \omega_{\rm b}, \, \omega_{\rm c}, \, \theta_s, \,  A_{\rm s}, \, n_s, \, \tau_{\rm reio}, 
\,   z_\dagger \right\},
\end{equation}
where the first six parameters are the baseline parameters of the standard $\Lambda$CDM model: $\omega_{\rm b}=\Omega_{\rm b} h^2$ and $\omega_{\rm c}=\Omega_{\rm c}h^2$ are the physical density parameters of baryons and cold dark matter today, respectively, $\theta_{\rm s}$ is the ratio of the sound horizon to the angular diameter distance at decoupling, $A_{\rm s}$ is the power of the primordial curvature perturbations at $k=0.05$ Mpc$^{-1}$, $n_{\rm s}$ is the power-law index of the scalar spectrum, and $\tau_{\rm reio}$ is the Thomson scattering optical depth due to reionization. We use uniform priors $\omega_b\in[0.018,0.024]$, $\omega_{\rm c}\in[0.10,0.14]$, $100\,\theta_{\rm s}\in[1.03,1.05]$, $\ln(10^{10}A_{\rm s})\in[3.0,3.18]$, $n_{\rm s}\in[0.9,1.1]$, and $\tau_{\rm reio}\in[0.04,0.125]$ for the common free parameters of model parameters and $z_\dagger\in[1,3]$ for the additional free parameter of $\Lambda_{\rm s}$CDM, which is determined in accordance with the discussions regarding $z_\dagger$ in Sec. \ref{sec:zdagger}.

In order to constrain the models, we use the latest \textit{Planck} CMB and BAO data: we use the recently released full \textit{Planck} (2018) \cite{Planck:2018vyg} CMB temperature and polarization data which consist of the low-$l$ temperature and polarization likelihoods at $l \leq 29$, temperature (TT) at $l \geq 30$, polarization (EE) power spectra, and cross correlation of temperature and  polarization (TE). The \textit{Planck} (2018) CMB lensing power spectrum likelihood \cite{Planck2018:GL} is also included. Along with the \textit{Planck} CMB data, we consider the high-precision Baryon Acoustic Oscillation measurements (BAO) at different redshifts up to $z=2.36$, viz., Ly-$\alpha$ DR14, BAO-Galaxy consensus, MGS and 6dFGS as presented in \cite{Alam:2016hwk,Blomqvist:2019rah,Ata:2017dya,Agathe:2019vsu,Beutler2011, Anderson:2013zyy}. It is worth noting that we include Ly-$\alpha$ measurements in our BAO compilation as they have a substantial impact on the parameters of $\Lambda_{\rm s}$CDM, whereas they have a minor impact on the parameters of $\Lambda$CDM, which is why they were excluded from the default BAO compilation by the \textit{Planck} (2018) Collaboration \cite{Planck:2018vyg}. 
We do not include BBN constraints on $\omega_{\rm b}$ so that we can compare the constraints on $\omega_{\rm b}$ predicted from our analysis for different models with those from BBN without bias. We have implemented the model in a modified version of the \texttt{CosmoMC} \cite{Lewis:2002ah} code to sample over the parameter space and produce posterior distributions; and used the \texttt{MCEvidence} \cite{MCEvidence} algorithm to compute the Bayesian evidence used to perform a model comparison through the Jeffreys' scale \cite{Vazquez:2011xa}. See Ref.~\cite{Padilla:2019mgi}, and references therein, for an extended review of the cosmological parameter inference and model selection procedure. We obtain the observational constraints on all of the parameters of the models---$\Lambda_{\rm s}$CDM, $\Lambda_{\rm s}$CDM+$z_\dagger=2.32$ (a particular case of $\Lambda_{\rm s}$CDM), and $\Lambda$CDM (for comparison purposes)---by using first only the CMB data and then the combined CMB+BAO data. 

\begin{figure}[t!]
\includegraphics[trim = 1mm  3mm 1mm 1mm, clip, width=0.38\textwidth]{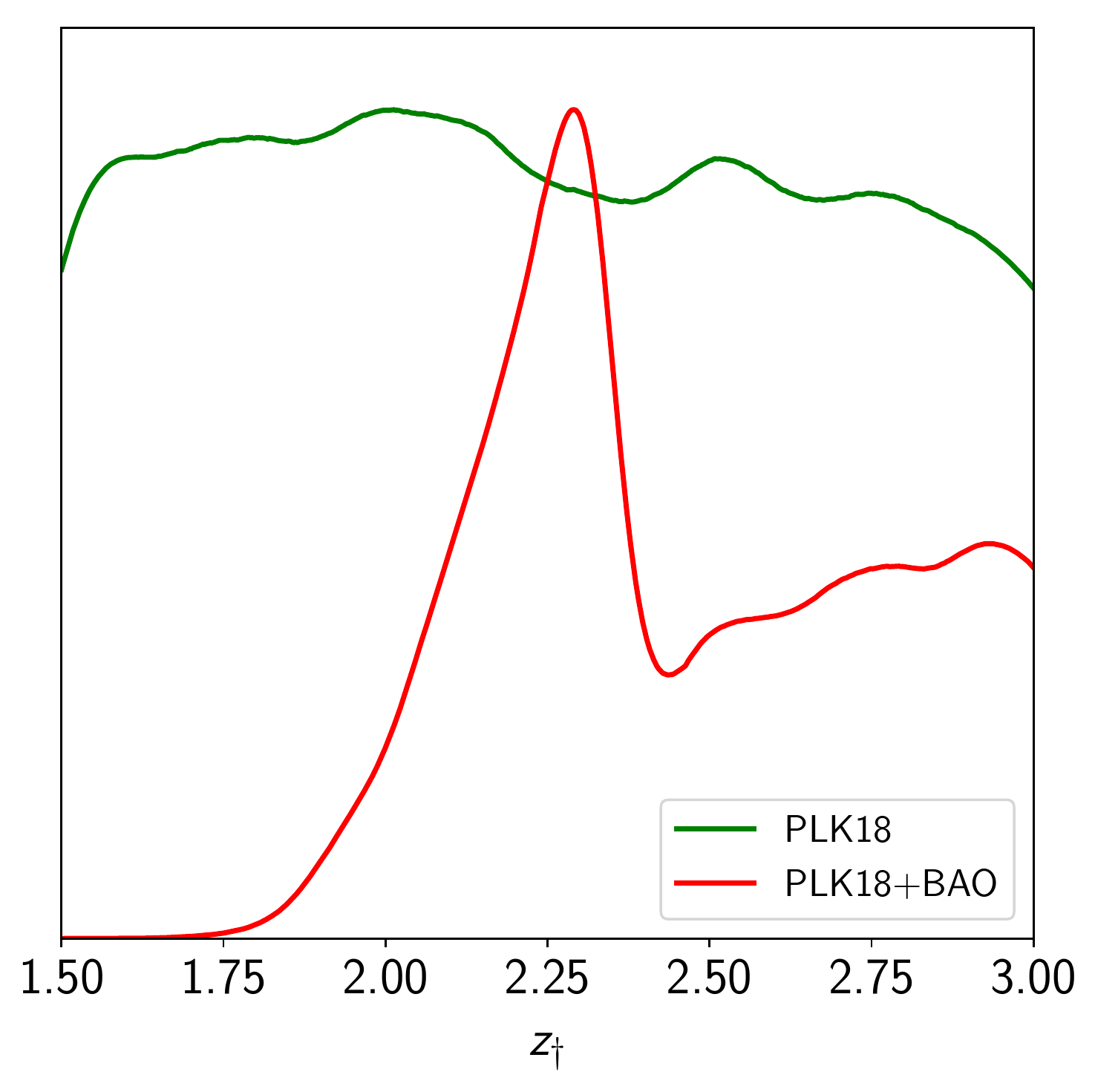}
\caption{One-dimensional marginalized distributions of the additional free parameter $z_\dagger$ of the $\Lambda_{\rm s}$CDM model. }
\label{1Dz}
\end{figure}

Table~\ref{tab:priors} displays the constraints at 68\% confidence level (CL) on the free parameters---$10^2\omega_{\rm b}$, $\omega_{\rm c}$, $100\,\theta_{\rm s}$, $\ln(10^{10}A_{\rm s})$, $n_{\rm s}$, $\tau_{\rm reio}$, and $z_\dagger$---as well as some derived parameters---the dust density parameter today $\Omega_{\rm m}$, the Hubble constant $H_0$, the amplitude of matter fluctuation on $8 h^{-1}$ Mpc comoving scale $\sigma_8$, and the combination $S_8\equiv\sigma_8\sqrt{\Omega_{\rm m}/0.3}$---from CMB and CMB+BAO data sets separately. We notice tight constraints on all of the model parameters from the combined CMB+BAO data, as expected. The additional parameter $z_\dagger$ in $\Lambda_{\rm s}$CDM is not constrained by the CMB data alone, as may also be seen from Fig.~\ref{1Dz} where the one-dimensional marginalized distributions of $z_\dagger$ are shown from the CMB and CMB+BAO data. 

In Fig.~\ref{1Dz}, we see that the one-dimensional marginalized distribution for $z_\dagger$ is quite flat for the CMB-only analysis (the green curve). The CMB data is insensitive to the value of $z_\dagger$ and cannot constrain it, as mentioned in Table~\ref{tab:priors}, because for any $z_\dagger\in[1.5,3]$ with $\omega_{\rm b}+\omega_{\rm c}\sim0.14$, there exists a $\Lambda_{\rm s0}$ value for which the comoving angular diameter distance to last scattering fits the CMB measurements. When the BAO data are included in the analysis (the red curve), however, the shape of the distribution changes dramatically, and we see a clear peak at $z_\dagger\approx2.3$. This is in line with the discussions in the previous section regarding the Ly-$\alpha$ and Galaxy BAO (SDSS DR14) data. We read off from Fig.~\ref{1Dz} that $z_\dagger$ must be larger than approximately $1.75$. The existence of a robust lower bound for $z_\dagger$ is no surprise, as we anticipated in the previous section from Fig.~\ref{fig:prelim_evo} that smaller $z_\dagger$ values correspond to higher tension with respect to the Galaxy BAO measurements. This behavior, in turn, decreases the probability of $z_\dagger$ for values smaller than $z_\dagger\approx2.3$ just before (in redshift) the redshift of the Ly-$\alpha$ measurements from $z\approx2.34$. On the other hand, we also see that there is a strong preference for $z_\dagger\lesssim 2.4$ since for these $z_\dagger$ values the $\Lambda_{\rm s}$CDM model has substantially better agreement with the Ly-$\alpha$ measurements, which is immediately lost for $z_\dagger\gtrsim 2.4$; just after (in redshift) the redshift of the Ly-$\alpha$ measurements from $z\approx2.34$, there is still a plateau-like tail for $z_\dagger\gtrsim2.4$ that is reminiscent of the green curve with the addition of a noticeable but insignificant trend towards larger $z_\dagger$ values. We refer the readers to Ref.~\cite{Akarsu:2019hmw} for a similar but more pronounced behavior caused by the Ly-$\alpha$ data (BOSS DR11) in gDE. 
Once $z_\dagger$ is restricted to this interval, the fit to the Ly-$\alpha$ data is essentially unaffected by the value of $z_\dagger$ and the data set is insensitive to $z_\dagger$, similar to the CMB-only analysis, except for the slight preference of the larger $z_\dagger$ values due to the presence of the Galaxy BAO data. In summary, the Ly-$\alpha$ data prefers $z_\dagger<2.34$ and the Galaxy BAO data pushes $z_\dagger$ to large values as much as possible; Fig.~\ref{1Dz} reflects the competition between the two results in the peak at $z_\dagger\approx2.3$. 
The asymmetric shape of the posterior for $z_\dagger$ that is not suitable to be approximated by a Gaussian or another standard distribution renders it not easily interpretable. 
For this reason, we also study a restriction of the $\Lambda_{\rm s}$CDM model denoted by ``$\Lambda_{\rm s}$CDM$+z_\dagger=2.32$" for which the only difference compared to $\Lambda_{\rm s}$CDM is that $z_\dagger$ is fixed to $2.32$, leaving 6 free parameters behind as in $\Lambda$CDM. The justification for our choice $z_\dagger=2.32$ is as follows. In Ref.~\cite{Akarsu:2019hmw}, it was the mean value of the constraints on $z_\dagger$ (denoted by $z_*$ there) both when $\lambda$ was free and was chosen with a large negative value making the gDE density behave like a step function imitating a sign-switching cosmological constant. Also, $z_\dagger=2.32$ is just slightly smaller than the redshift of the Ly-$\alpha$ measurements $z\approx2.34$, and is supposed to provide better agreement with the Ly-$\alpha$ measurements; this value is also very close to both the peak and the mean of the red posterior in Fig.~\ref{1Dz}. The constraints on the $\Lambda_{\rm s}$CDM$+z_\dagger=2.32$ model parameters are given in Table~\ref{tab:priors}.

\begin{figure}[t!]\centering
\includegraphics[trim = 1mm  3mm 1mm 1mm, clip, width=0.38\textwidth]{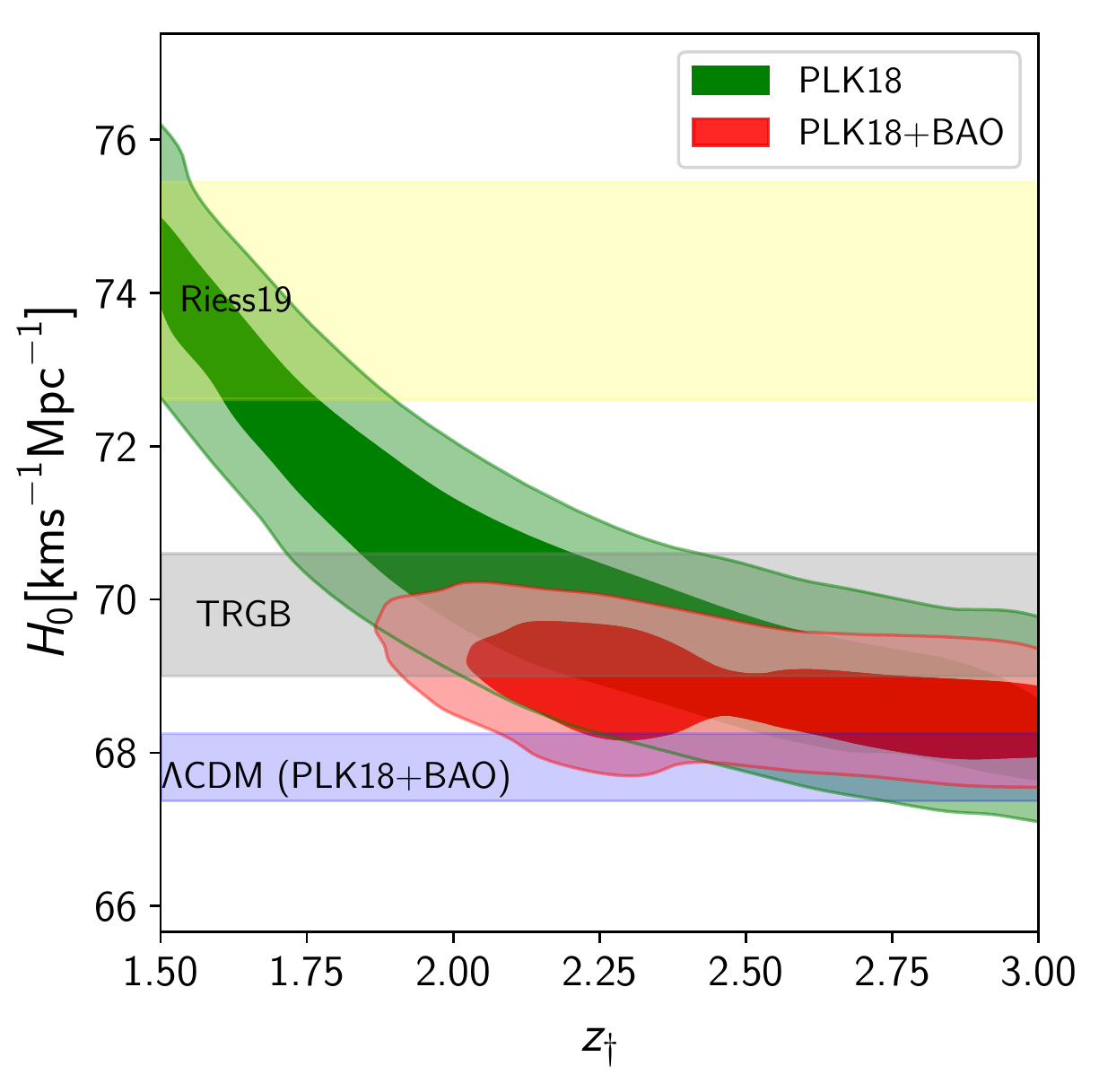} 
\caption{Two-dimensional (68\% and 95\% C.L.) marginalized distributions of $H_0$ versus $z_\dagger$ for the $\Lambda_{\rm s}$CDM model, showing a negative correlation between the two parameters, which implies that smaller values of $z_\dagger$ correspond to larger values of $H_0$. } 
\label{fig:hzstar}
 \end{figure}

\begin{figure}[t!]\centering
\includegraphics[trim = 1mm  3mm 1mm 1mm, clip, width=0.38\textwidth]{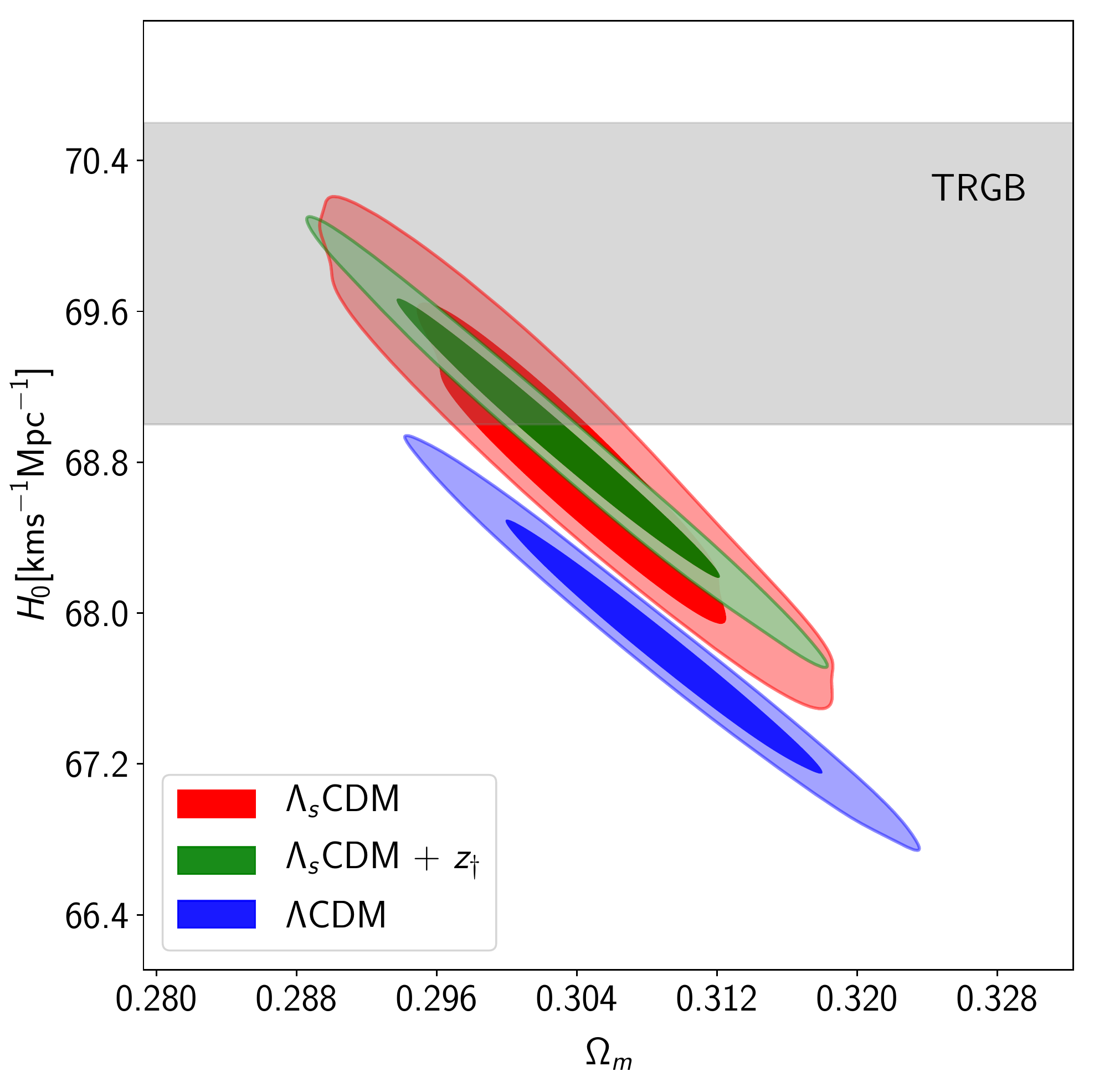} 
\caption{Two-dimensional (68\% and 95\% C.L.) marginalized distributions of $H_0$ versus $\Omega_{\rm m}$ from CMB+BAO data, showing how the $H_0$ tension is relaxed in the $\Lambda_{\rm s}$CDM model compared to the $\Lambda$CDM model wherein the horizontal gray band is for the model-independent TRGB $H_0$ measurement $H_0= 69.8 \pm 0.8\, {\rm km\,s}^{-1}{\rm Mpc}^{-1}$ \cite{Freedman:2019jwv}.} 
\label{fig:H0_vs_Omegam}
 \end{figure}

In Fig.~\ref{fig:hzstar} we show the two-dimensional (68\% and 95\% C.L.) marginalized distributions of $H_0$ versus $z_\dagger$ from the CMB-only data set (green contours), and the combined CMB+BAO data set (red contours). We notice a negative correlation between these two parameters, as expected (see Sec.~\ref{sec:zdagger}). Since $z_\dagger$ is not constrained by the CMB-only data set, the green contours scan the whole range of $z_\dagger$; also, as we anticipated from Fig.~\ref{fig:zstar_vs_H0}, they encompass even the largest model-independent measurements of $H_0$ up to $\sim74$ km s${}^{-1}$ Mpc${}^{-1}$. Due to their strong correlation, the constraints on $z_\dagger$ are also directly reflected in $H_0$, and the exclusion of low  $z_\dagger$ values by the Galaxy BAO data corresponds to the exclusion of the highest $H_0$ values. For the CMB+BAO data set, $2.15<z_\dagger<2.73$ at $68\%$ C.L., as can be read from Table~\ref{tab:priors}, and this prevents the red contours from containing $H_0$ values as high as the green one, yet $H_0= 68.82 \pm 0.55\, {\rm km\,s}^{-1}{\rm Mpc}^{-1}$ ($H_0= 68.91 \pm 0.48\, {\rm km\,s}^{-1}{\rm Mpc}^{-1}$ for the $\Lambda_{\rm s}$CDM+$z_\dagger=2.32$) at $68\%$ C.L., is larger than $H_0= 67.81 \pm 0.4\, {\rm km\,s}^{-1}{\rm Mpc}^{-1}$ ($68\%$ C.L.) of the $\Lambda$CDM prediction, and in good agreement with the model-independent TRGB measurement $H_0= 69.8 \pm 0.8\, {\rm km\,s}^{-1}{\rm Mpc}^{-1}$ ($68\%$ C.L.) \cite{Freedman:2019jwv}. Since the impact of the sign switch feature becomes less effective for larger $z_\dagger$ values, both contours approach the $\Lambda$CDM interval of $H_0$ for large $z_\dagger$, but the error margin is larger for $\Lambda_{\rm s}$CDM due to the additional errors contributed by the uncertainty of the extra free parameter $z_\dagger$. Complimentary to the discussion in this paragraph, in Fig.~\ref{fig:H0_vs_Omegam} we show the two-dimensional (68\% and 95\% C.L.) marginalized distributions of $H_0$ versus $\Omega_{\rm m}$ from CMB+BAO data, which shows how the $H_0$ tension is relaxed in $\Lambda_{\rm s}$CDM compared to $\Lambda$CDM. There is a negative correlation between $H_0$ and $\Omega_{\rm m}$ for all three models. $\Lambda_{\rm s}$CDM does not overlap with $\Lambda$CDM even at $95\%$ C.L.; this separation is even more pronounced when the $z_\dagger=2.32$ restriction is considered. Unsurprisingly, $\Lambda_{\rm s}$CDM$+z_\dagger=2.32$ is contained within $\Lambda_{\rm s}$CDM and is tightly constrained just like $\Lambda$CDM which has the same number of free parameters.

\begin{figure}
     \centering
     \includegraphics[width=0.48\textwidth]{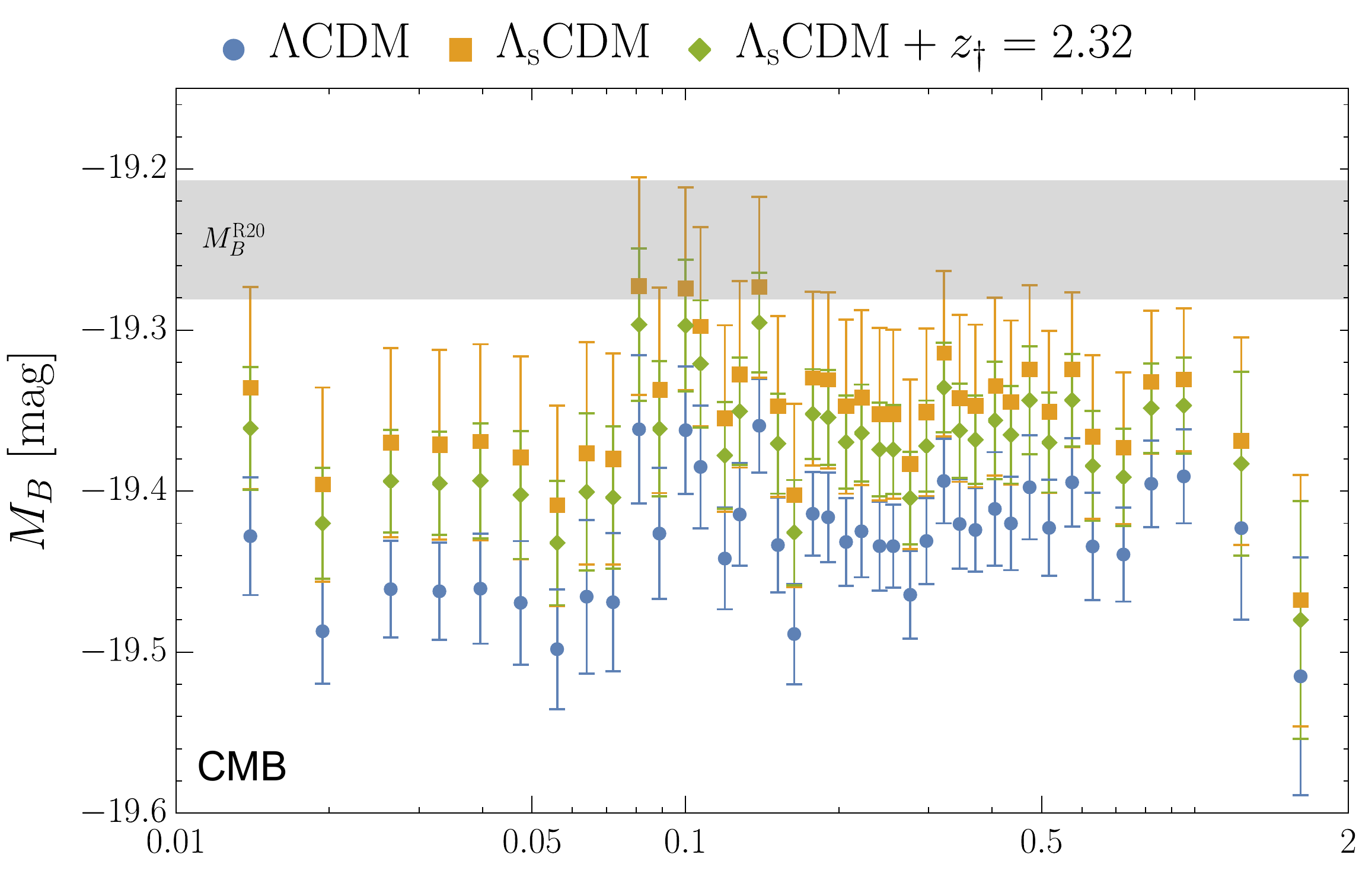}
     \includegraphics[width=0.48\textwidth]{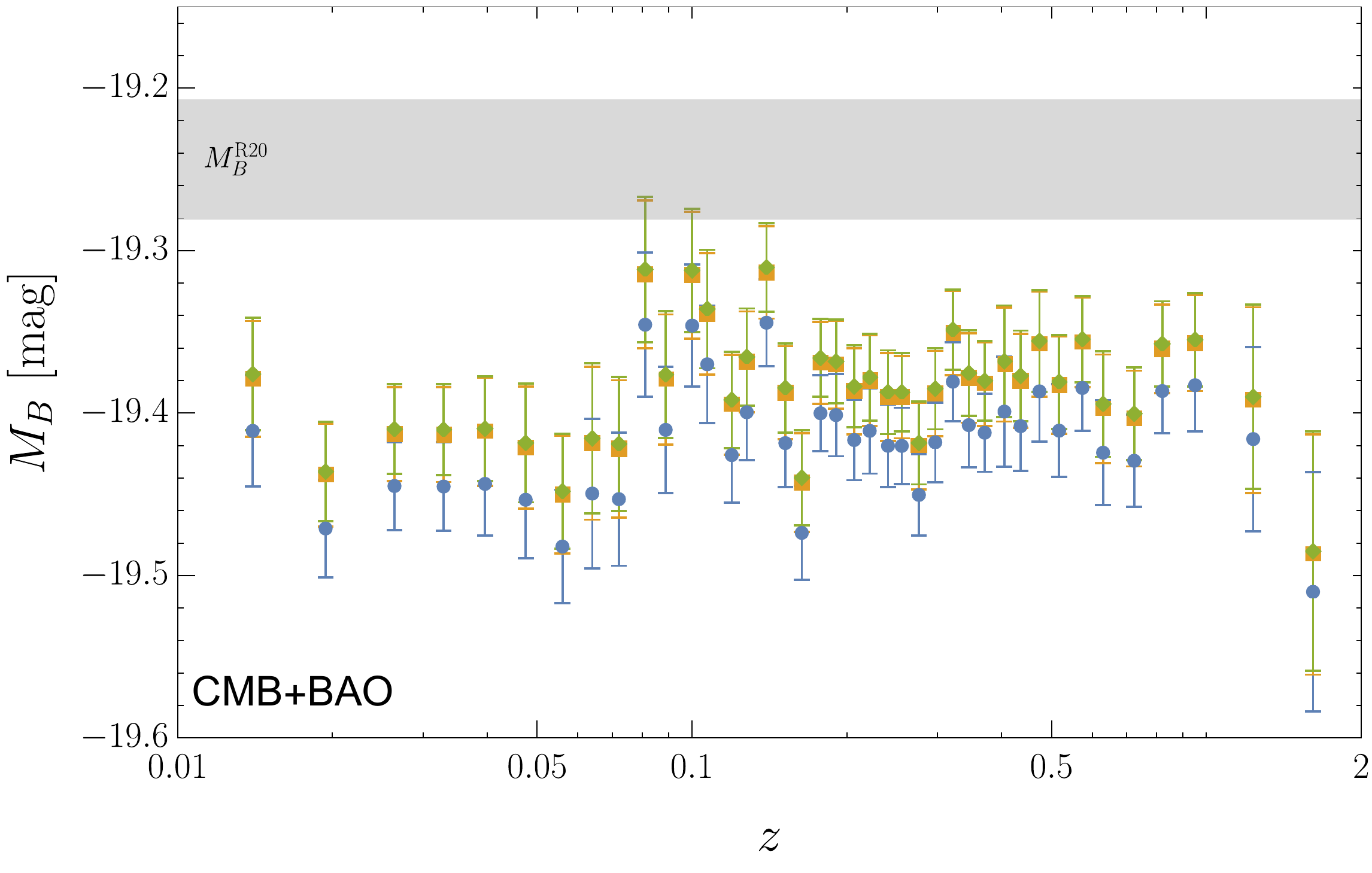}
     \caption{Observational counterpart of Fig.~\ref{fig:prelimMB} for the CMB-only (top panel) and combined CMB+BAO (bottom panel) analyses. The constraints on the absolute magnitudes ($M_{B,i}$) are obtained from $M_{B,i}=m_{B,i}-\mu(z_i)$ by using the apparent magnitudes ($m_{B,i}$) of the binned Pantheon SnIa sample \cite{Scolnic:2017caz} and the constraints we obtained at 68\% C.L. on the distance modulus values $\mu(z_i)$ for the corresponding SnIa data points.}
     \label{fig:obsMB}
 \end{figure}

We have discussed in Sec. \ref{sec:zdagger} that, within the $\Lambda_{\rm s}$CDM model, the amelioration of the SH0ES $H_0$ tension is accompanied by an amelioration of the $M_B$ tension respecting the internal consistency of the SH0ES measurements of these parameters. We have shown with a preliminary analysis that $M_{B,i}$ values calculated by subtracting the distance modulus from the apparent magnitudes of the binned Pantheon sample \cite{Scolnic:2017caz} should be greater for $\Lambda_{\rm s}$CDM compared to the standard model. In this section, we do the same $M_{B,i}$ calculations, but now we compute the distance modulus values directly from our data analysis; indeed, we see in Fig.~\ref{fig:obsMB} (the observational counterpart of Fig.~\ref{fig:prelimMB}) that the $\Lambda_{\rm s}$CDM models result in $M_{B,i}$ values that are systematically higher than those of $\Lambda$CDM (as they do for $H_0$ values) and have better agreement with the $M_B^{\rm R20}$ value (as they do with local measurements of $H_0$). For the CMB-only analysis in the top panel, the unrestricted $\Lambda_{\rm s}$CDM, which has the highest $H_0$ value agreeing the best with the SH0ES value, has also the best agreement with the $M_B^{\rm R20}$ value among the three models. When BAO is included in the data set, the restricted $\Lambda_{\rm s}$CDM, compared to the other two models, has the better agreement with the SH0ES $H_0$ value and thus (as seen from the bottom panel of Fig.~\ref{fig:obsMB}) also with $M_B^{\rm R20}$. $\Lambda$CDM, on the other hand, performs substantially worse for both the CMB-only and the combined CMB+BAO analyses. As the $M_B$ and SH0ES $H_0$ tensions are almost equivalent for $\Lambda_{\rm s}$CDM, just like they are for $\Lambda$CDM, the Galaxy BAO data (which effectively puts an upper bound on the $H_0$ values $\Lambda_{\rm s}$CDM can achieve), in parallel, also puts an upper bound on its $M_{B,i}$ predictions, limiting the success of the model in alleviating these tensions.

We see that there are certain distinctions between the CMB and the CMB+BAO analyses when parameters related to matter densities are considered. As seen from Table~\ref{tab:priors}, the CMB-only analysis puts very similar constraints (within $\sim1\sigma$ of each other) on $\omega_{\rm b}$, $\omega_{\rm c}$, and hence $\omega_{\rm m}\equiv\omega_{\rm b}+\omega_{\rm c}$ for all three models, while the constraints on $\Omega_{\rm m}$ vary among the models. In this case, all three $\omega_{\rm b}$ values present similar discrepancies compared to the BBN constraint $10^2\omega_{\rm b}=2.166\pm0.019$ (namely, $10^2\omega_{\rm b}=2.166\pm0.015\pm0.011$, where the first error term is due to the uncertainty in the measurement of the primordial deuterium abundance and the second error term is due to the uncertainty in the BBN calculations) \cite{Cooke:2017cwo}.
Note that this BBN constraint is based on the $d(p,\gamma)^3\rm He$ reaction rate computed in Ref.~\cite{Marcucci:2015yla}. Interestingly, including the BAO data in the analysis puts similar constraints on $\Omega_{\rm m}$ (within $\sim1\sigma$ of each other) for all three models while letting $\omega_{\rm b}$ and $\omega_{\rm c}$ vary among the models. This has some important consequences. First, the BAO data pull $\omega_{\rm m}=\Omega_{\rm m}h^2$ towards smaller values for $\Lambda$CDM but towards greater values for both of the $\Lambda_{\rm s}$CDM models; given the similar $\Omega_{\rm m}$ values for all three, this results in higher $H_0$ values for the $\Lambda_{\rm s}$CDM models compared to $\Lambda$CDM. Second, $\omega_{\rm b}$ follows a reverse trend for all models compared to $\omega_{\rm m}$, i.e., the BAO data pull $\omega_{\rm b}$ towards greater values for $\Lambda$CDM while it is pulled towards smaller values for both of the $\Lambda_{\rm s}$CDM models. Thus, with the inclusion of the BAO data, the discrepancy with the BBN constraint for $\omega_{\rm b}$ worsens for $\Lambda$CDM while relaxes for the $\Lambda_{\rm s}$CDM models. We wonder if this amelioration for the $\Lambda_{\rm s}$CDM model could be improved if the Galaxy BAO data were not present in the analysis. Note that in Ref.~\cite{Cooke:2017cwo} they also presented the value $10^2\omega_{\rm b}=2.235\pm0.037$ (namely, $10^2\omega_{\rm b}=2.235\pm0.016\pm0.033$) when the empirical $d(p,\gamma)^3\rm He$ reaction rate in Ref.~\cite{Adelberger:2010qa} was used; even in this case, the $\Lambda_{\rm s}$CDM models are in better agreement with the BBN constraint for $\omega_{\rm b}$ when the CMB+BAO data set is considered.

\begin{figure*}[ht!]\centering
\includegraphics[trim = 1mm  5mm 22mm 5mm, clip, width=7.5cm, height=4.cm]{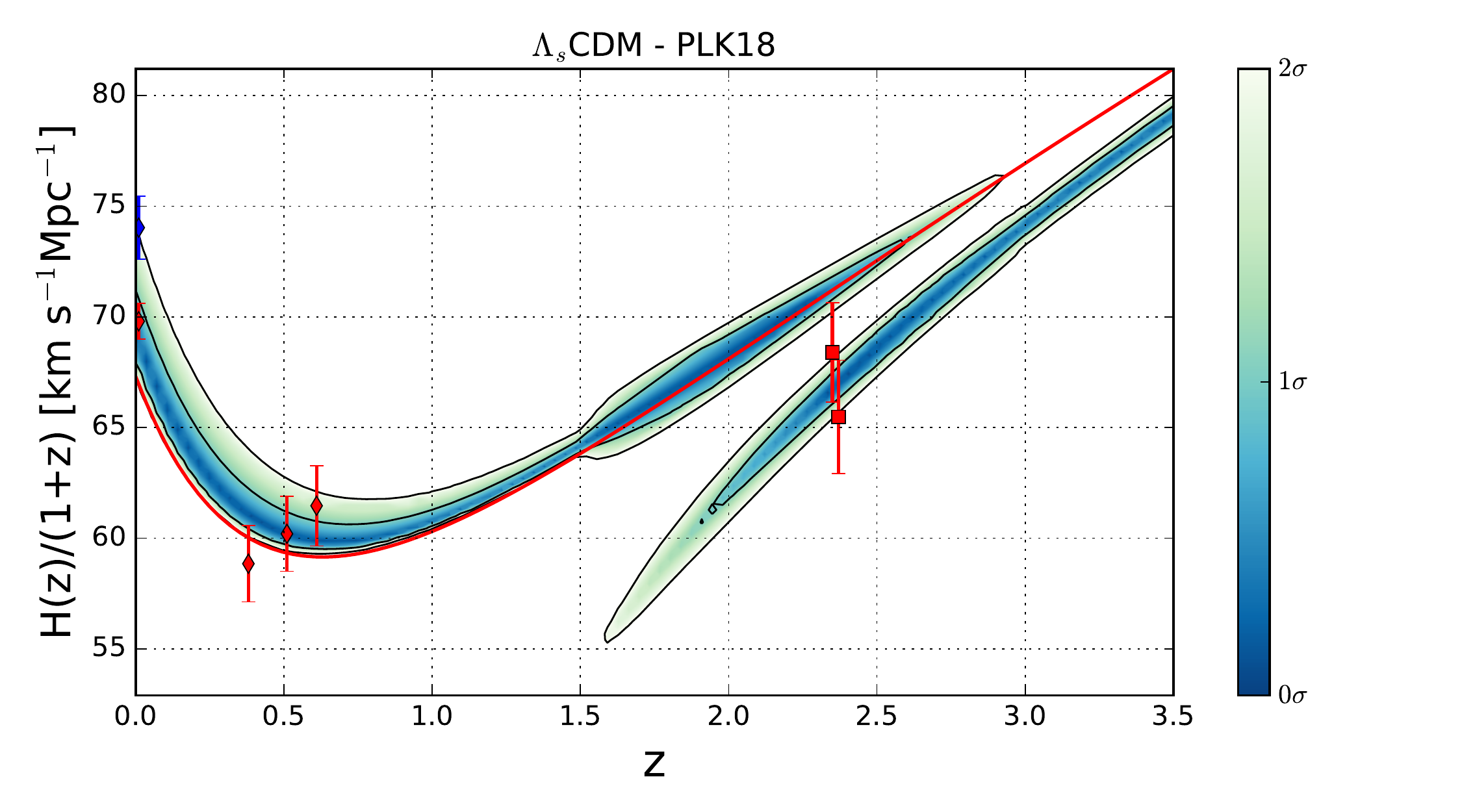} 
\includegraphics[trim = 1mm  5mm 22mm 5mm, clip, width=7.5cm, height=4.cm]{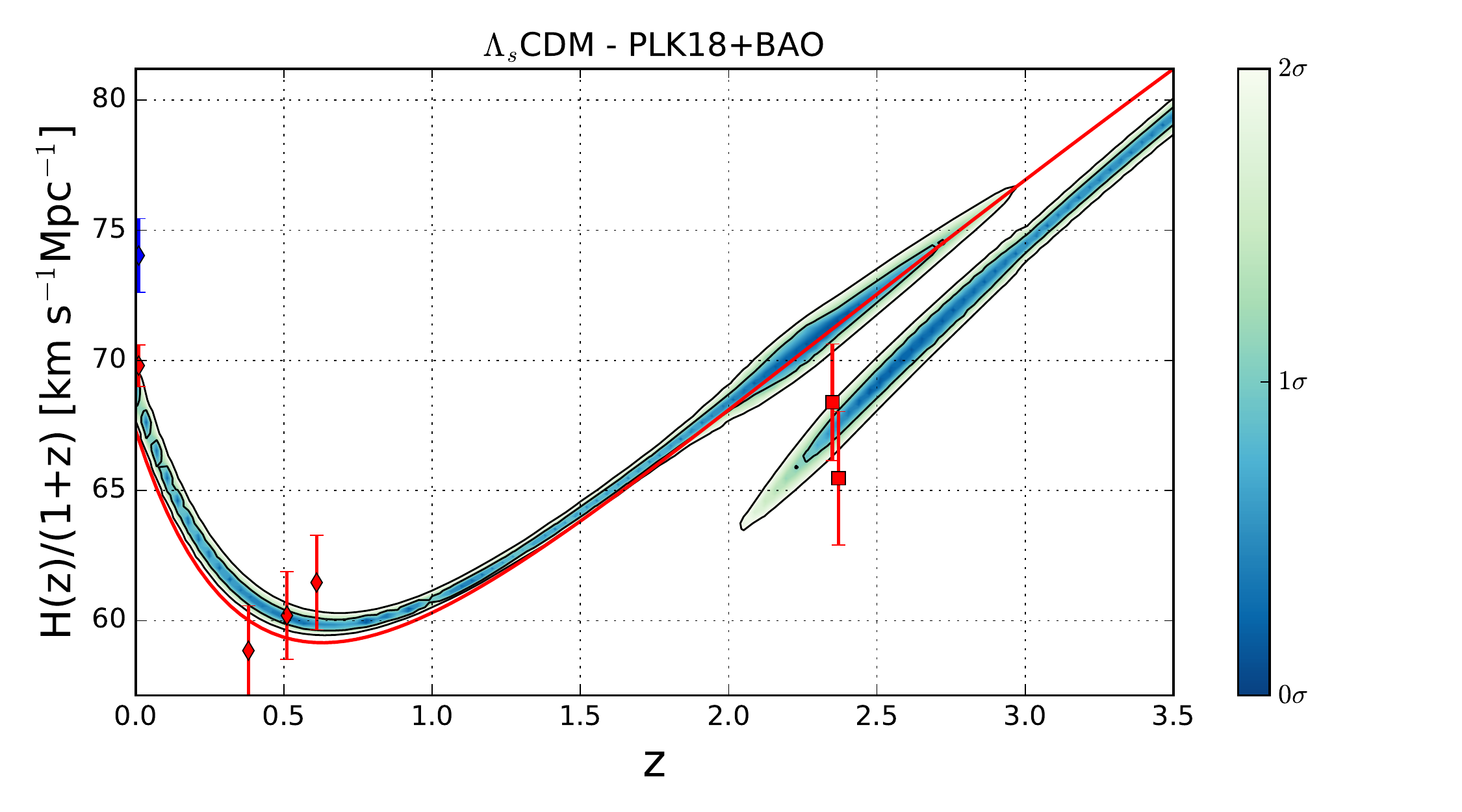} \\

\includegraphics[trim = 1mm  5mm 22mm 5mm, clip, width=7.5cm, height=4.cm]{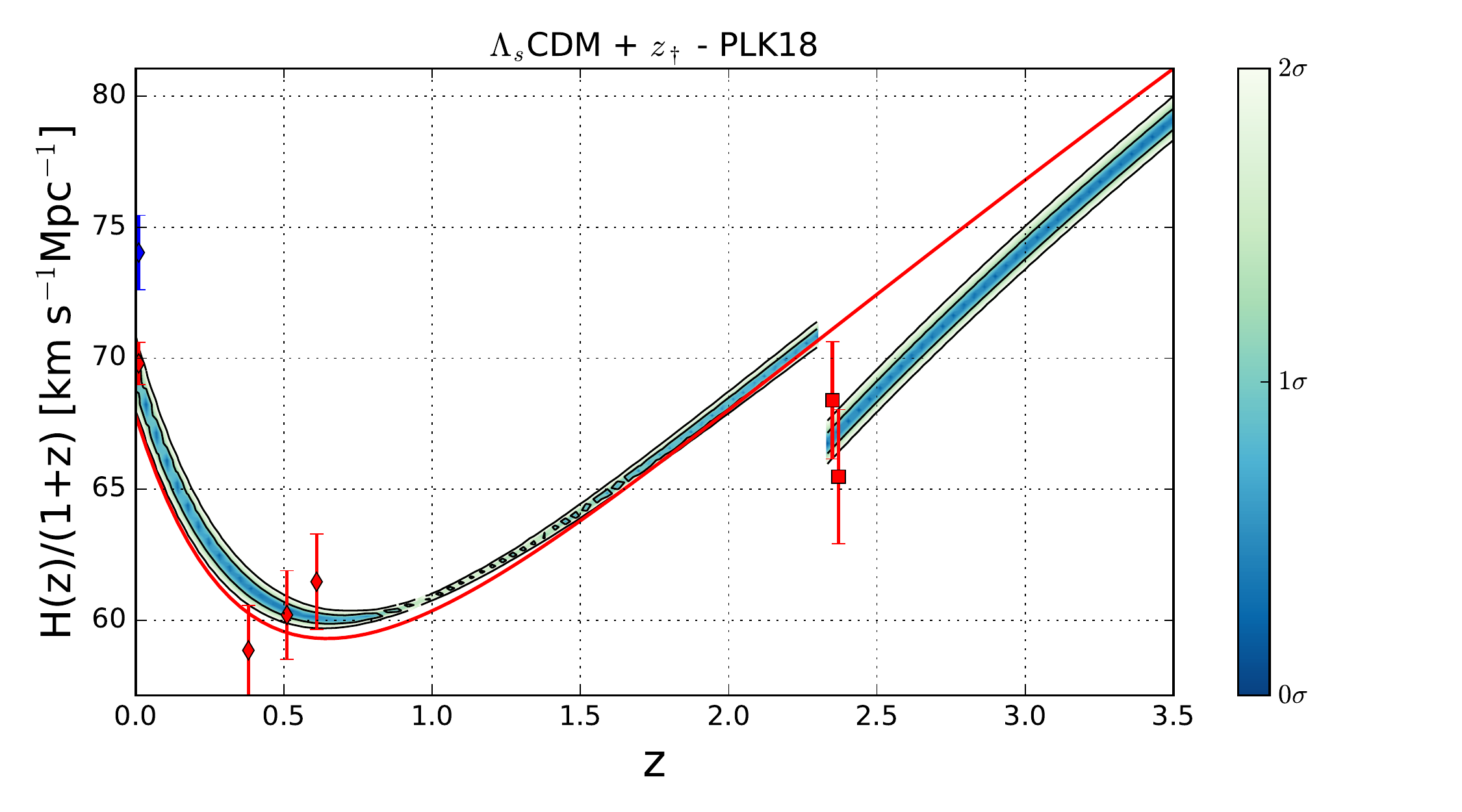} 
\includegraphics[trim = 1mm  5mm 22mm 5mm, clip, width=7.5cm, height=4.cm]{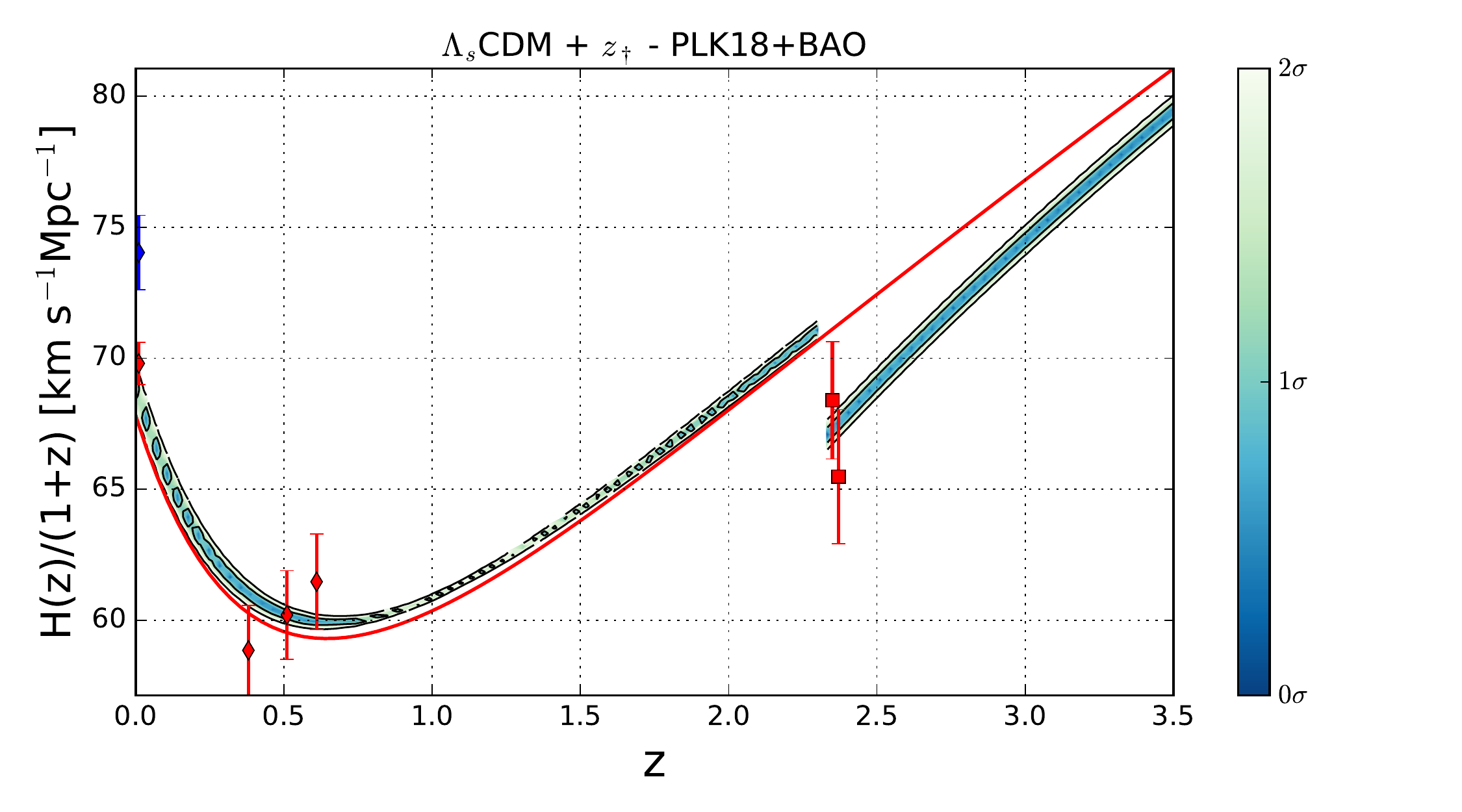} 

\caption{$H(z)/(1+z)$ versus $z$ with 68\% and 95\% error regions in case of CMB (left panel) and CMB+BAO (right panel) data, showing how the Ly-$\alpha$ data tension is relaxed in the $\Lambda_{\rm s}$CDM model compared to the $\Lambda$CDM model, wherein the red curve stands for the $\Lambda$CDM model corresponding to the mean values of the parameters.  We show the observational $H(z)$ values (error bars): $H_0=69.8\pm0.8\,{\rm km\,s}^{-1}{\rm Mpc}^{-1}$ from the TRGB $H_0$ \cite{Freedman:2019jwv}, $H_0=74.03\pm1.42\, {\rm km\,s}^{-1}{\rm Mpc}^{-1}$ from the Cepheid measurement $H_0$ \cite{Riess:2019cxk}, BAO Galaxy consensus (from $z_{\rm eff}=0.38,\,0.51,\,0.61$), and Ly-$\alpha$ DR14 (from $z_{\rm eff}=2.34,\,2.35$) \cite{Blomqvist:2019rah,Anderson:2013zyy}.} 
\label{fig:evolution}
 \end{figure*}
 
\begin{figure*}[t!]\centering
\includegraphics[trim = 0mm  0mm 0mm 0mm, clip, width=7.5cm, height=4.cm]{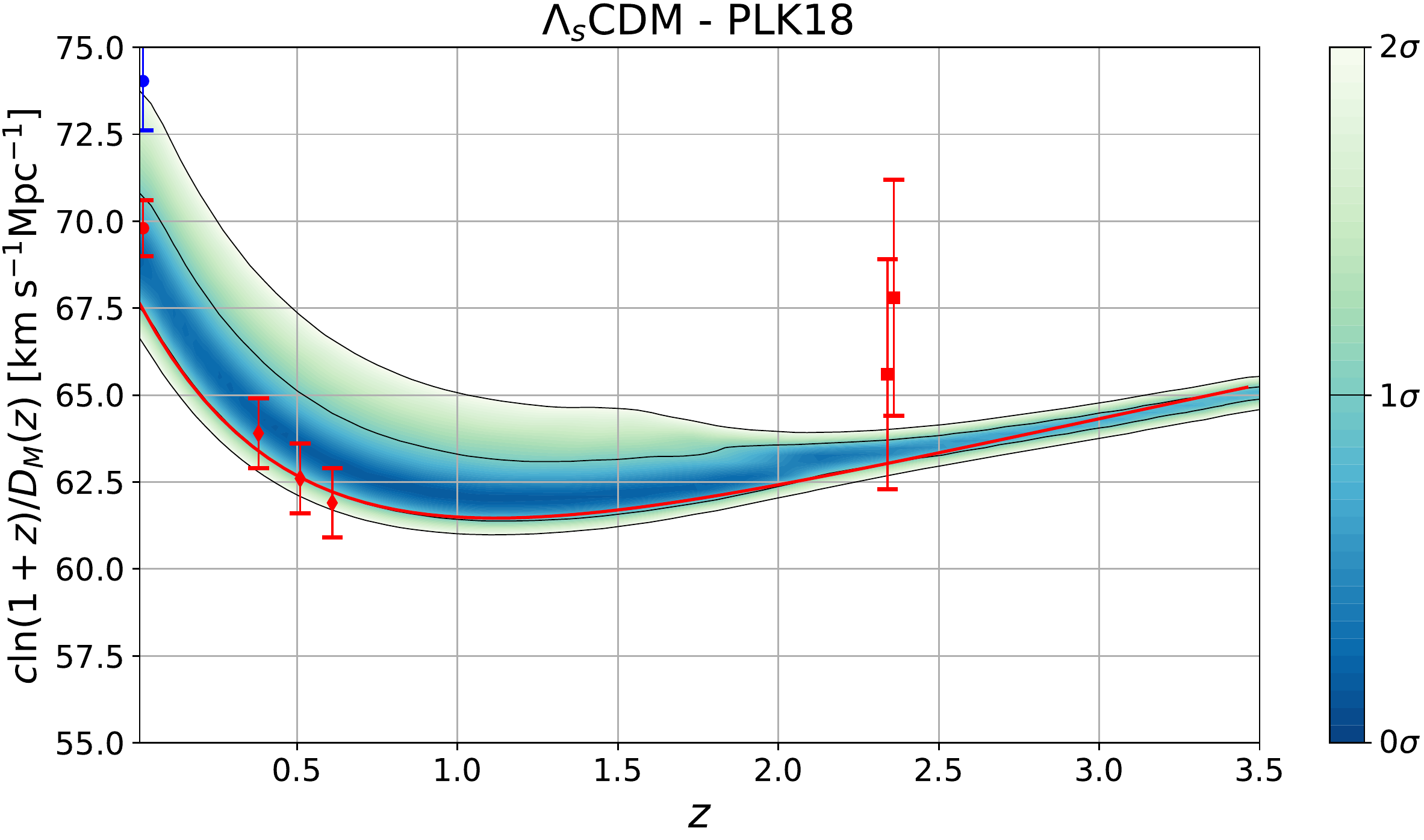} 
\includegraphics[trim = 0mm  0mm 0mm 0mm, clip, width=7.5cm, height=4.cm]{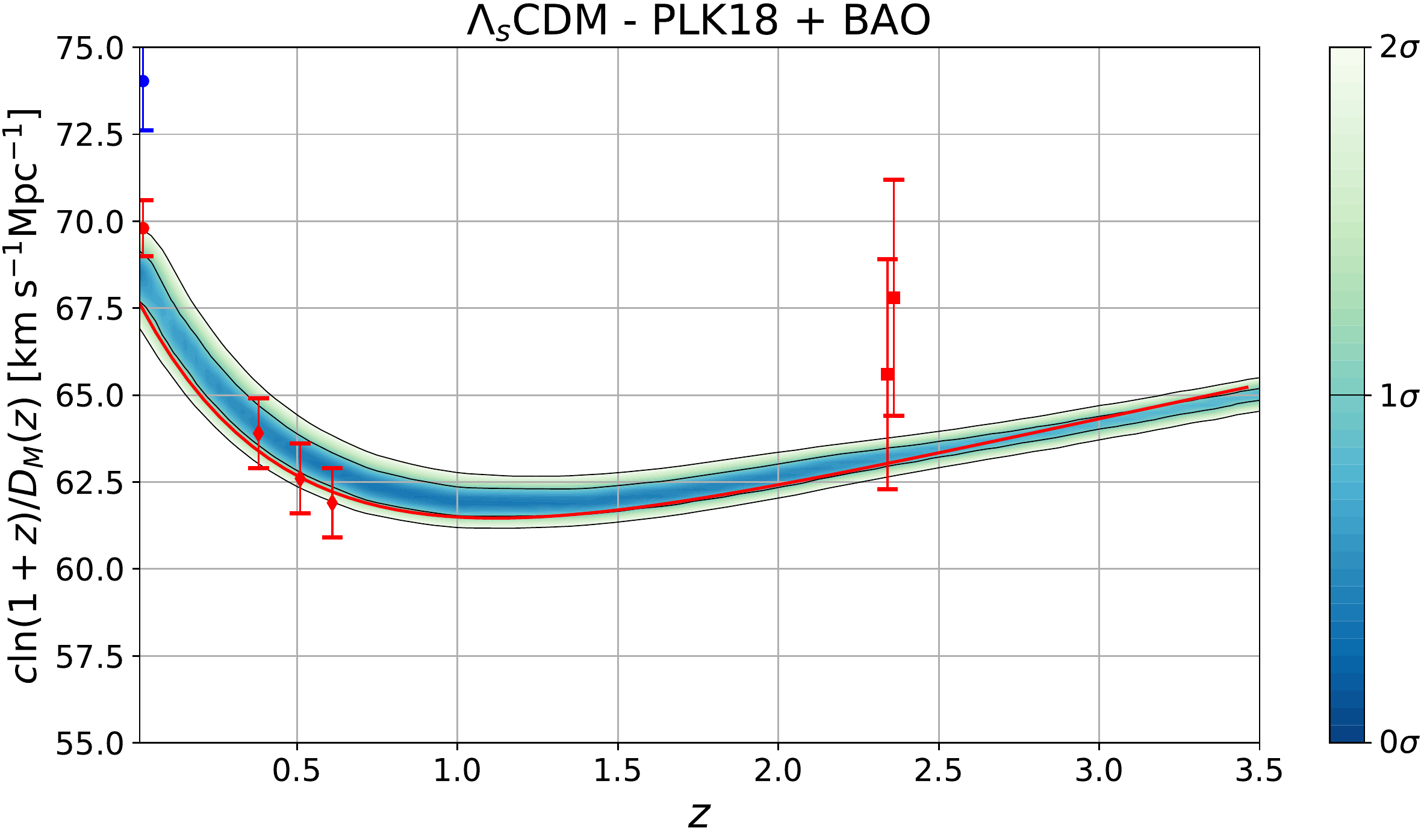} \\

\includegraphics[trim = 0mm  0mm 0mm 0mm, clip, width=7.5cm, height=4.cm]{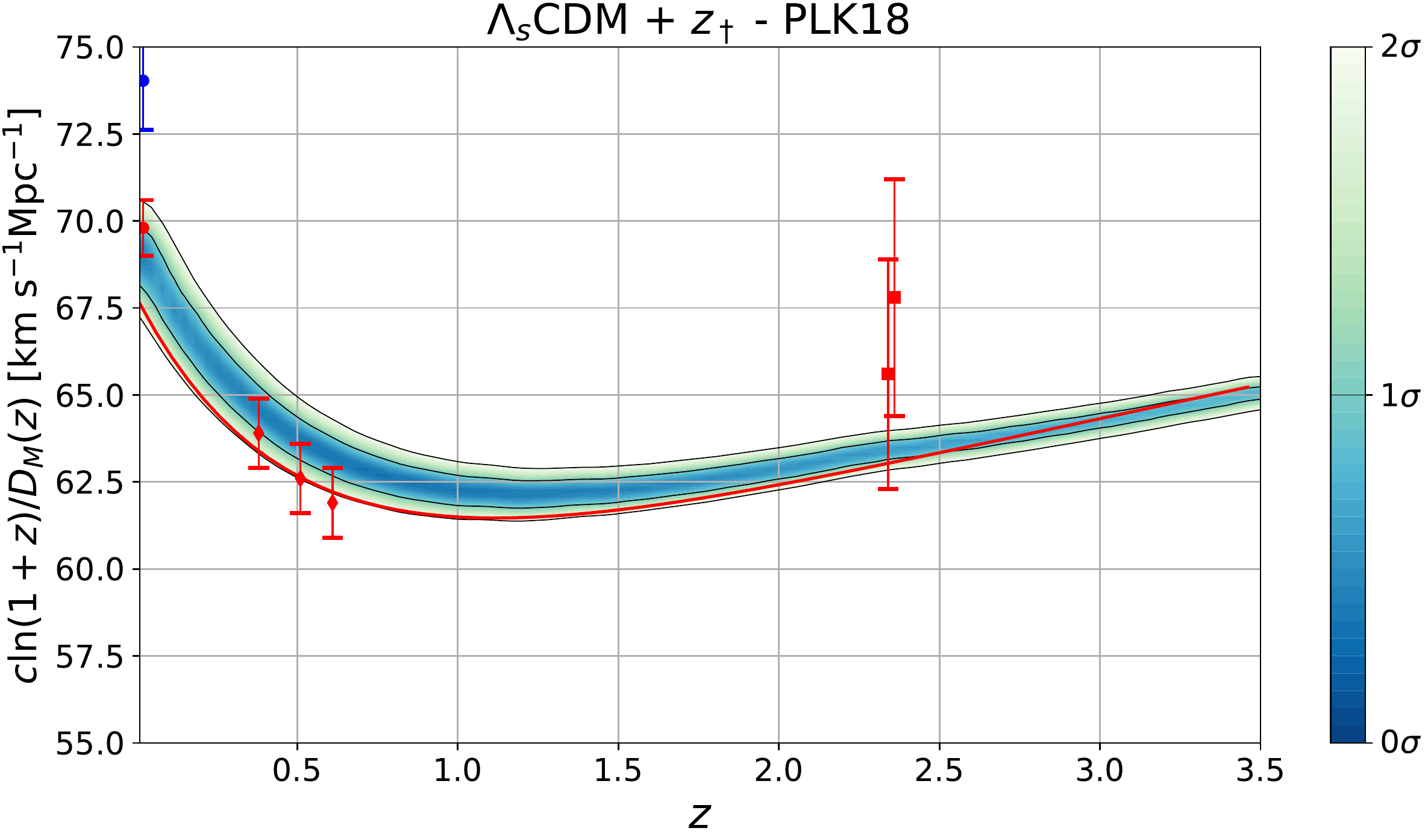} 
\includegraphics[trim = 0mm  0mm 0mm 0mm, clip, width=7.5cm, height=4.cm]{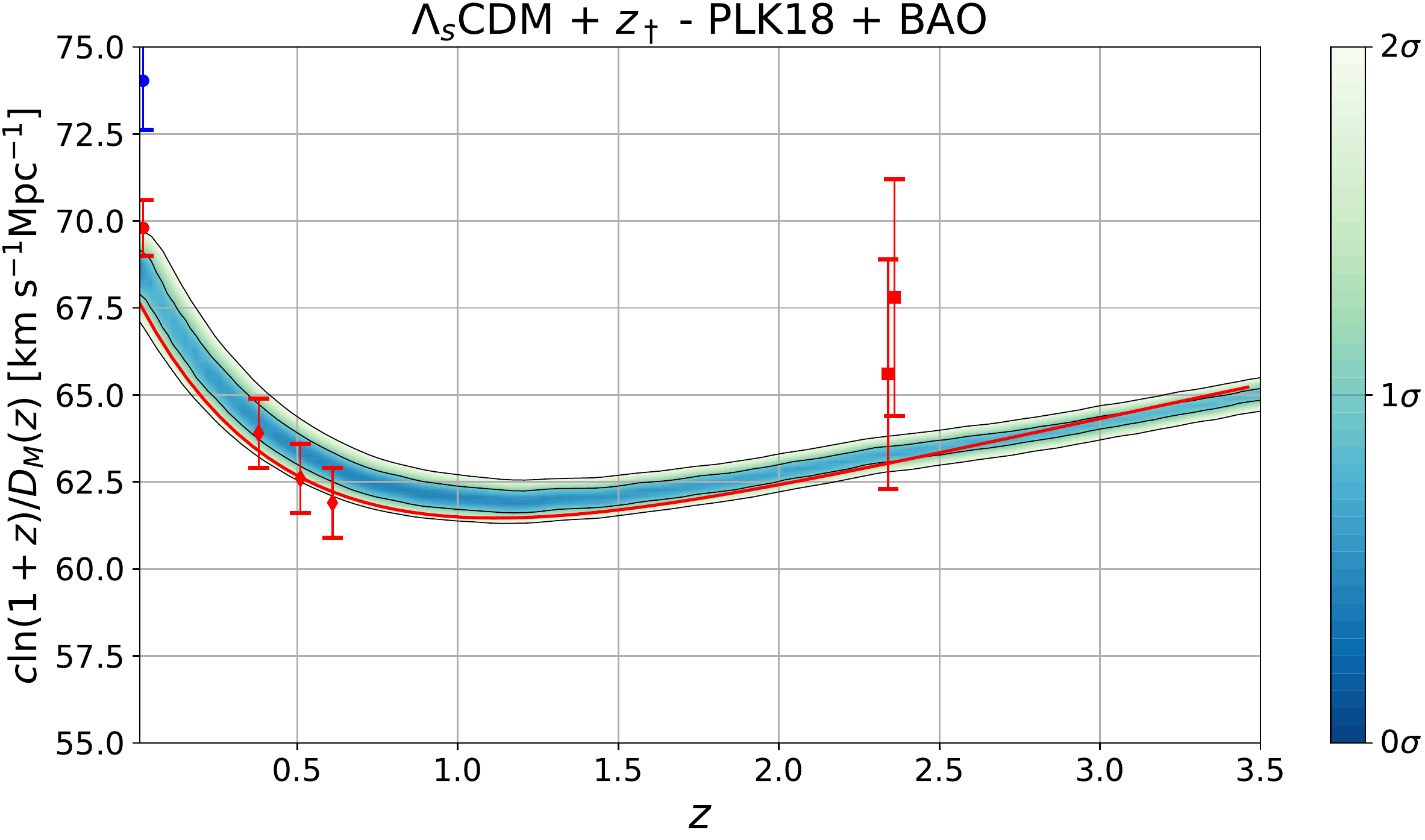} 

\caption{$c\ln(1+z)1/D_M(z)\equiv \mathcal{D}(z)$ versus $z$ with 68\% and 95\% error regions in case of CMB (left panel) and CMB+BAO (right panel) data. We show the observational $H(z)$ values (error bars): $H_0=69.8\pm0.8\,{\rm km\,s}^{-1}{\rm Mpc}^{-1}$ from the TRGB $H_0$ \cite{Freedman:2019jwv}, $H_0=74.03\pm1.42\, {\rm km\,s}^{-1}{\rm Mpc}^{-1}$ from the Cepheid measurement $H_0$ \cite{Riess:2019cxk}, BAO Galaxy consensus (from $z_{\rm eff}=0.38,\,0.51,\,0.61$), and Ly-$\alpha$ DR14 (from $z_{\rm eff}=2.34,\,2.35$) \cite{Blomqvist:2019rah,Anderson:2013zyy}.
} 
\label{fig:distance}
 \end{figure*}

In Fig.~\ref{fig:evolution} (the observational counterpart of the top panel of Fig.~\ref{fig:prelim_evo}), obtained by using the \texttt{fgivenx} \texttt{PYTHON} package \cite{handley19}, we show $H(z)/(1+z)$ versus $z$ with probability regions up to 95\% C.L. (the darker implies more probable, as shown in the color bar) for CMB (left panel) and CMB+BAO (right panel) data sets, showing how the discrepancy with the Ly-$\alpha$ measurements disappears completely in $\Lambda_{\rm s}$CDM compared to the $\Lambda$CDM model wherein we show the observational $H(z)$ values $H_0=69.8\pm0.8\,{\rm km\,s}^{-1}{\rm Mpc}^{-1}$ from the TRGB $H_0$ \cite{Freedman:2019jwv}, $H_0=74.03\pm1.42\, {\rm km\,s}^{-1}{\rm Mpc}^{-1}$ from the local measurements using Cepheid calibrators
\cite{Riess:2019cxk}, BAO Galaxy consensus (from effective redshifts $z_{\rm eff}=0.38,\,0.51,\,0.61$) and Ly-$\alpha$ DR14 (from effective redshifts $z_{\rm eff}=2.34,\,2.35$) \cite{Blomqvist:2019rah,Anderson:2013zyy}. The inclusion of the BAO data in the analysis substantially tightens the constraints on $H(z)$ for both models. This also lowers the maximum $H_0$ value contained within the $2\sigma$ contours for both models and this effect is more pronounced in the unrestricted $\Lambda_{\rm s}$CDM due to the truncation of the smaller $z_\dagger$ values by the Galaxy BAO data. 
Indeed, while the unrestricted model is in partial agreement with the $H_0$ value from the Cepheid measurements for the CMB-only analysis, for the CMB+BAO data set a significant tension appears, but the model is still in very good agreement with the $H_0$ value from the TRGB measurements. For $z\lesssim 2.3$, the mean $H(z)$ curve of $\Lambda$CDM is below both of the $\Lambda_{\rm s}$CDM models and for $z\lesssim1.5$ it (including $H_0$) is even excluded in the $95\%$ C.L.. 
For $z\gtrsim3$, both $\Lambda_{\rm s}$CDM models strongly exclude the mean $H(z)$ curve of $\Lambda$CDM  by preferring lower values, but the unrestricted $\Lambda_{\rm s}$CDM has an interval of compatibility with $\Lambda$CDM for $2.3\lesssim z\lesssim 3$ at the cost of losing its improved fit to the Ly-$\alpha$ data. It is not clear from this figure how $\Lambda_{\rm s}$CDM, compared to the $\Lambda$CDM model, responds to the Galaxy BAO data; as we have discussed in the previous sections, the opposition of the Galaxy BAO data to the smaller $z_\dagger$ values is based on the comoving angular diameter distance $D_M(z)$ measurements.

In Fig.~\ref{fig:distance} (the observational counterpart of the bottom panel of Fig.~\ref{fig:prelim_evo}) we show $\mathcal{D}(z)\equiv c\ln(1+z)/D_M(z)$ versus $z$ with probability regions up to 95\% C.L. for both $\Lambda_{\rm s}$CDM models, and the mean $\mathcal{D}(z)$ curve for the $\Lambda$CDM model. We see from the top left panel that the distribution for the unrestricted $\Lambda_{\rm s}$CDM for the CMB-only analysis diffuses to substantially higher values compared to $\Lambda$CDM, and, is almost always above $\Lambda$CDM; in fact, the mean curve for $\Lambda$CDM acts almost as a lower bound for the $2\sigma$ contours of $\Lambda_{\rm s}$CDM. Note that the lowest parts of the contours correspond to the highest redshifts for the sign switch, i.e., to $z_\dagger\sim3$. This behavior of elevated $\mathcal{D}(z)$ translates into the preference for higher $H_0$ values at $z=0$ in the presence of the sign switch.
 When the BAO data is included in the analysis, the posterior changes very slightly around the Ly-$\alpha$ data and the improved agreement is present for both data sets; in contrast, the inclusion of the BAO data strictly reduces the spread of the distribution at lower $z$ values and excludes $H_0\gtrsim70\,{\rm km\,s}^{-1}{\rm Mpc}^{-1}$ in the $2\sigma$ CL, but the mean curve for $\Lambda$CDM still acts almost as a lower bound. This shows that higher $\mathcal{D}(z)$ values compared to $\Lambda$CDM are characteristic of the $\Lambda_{\rm s}$CDM model. 
 For the $\Lambda_{\rm s}$CDM+$z_\dagger=2.32$ model, the story is very similar but less emphasized. The spread of the posterior is thinner due to the absence of the uncertainty contributed by $z_\dagger$, and including the BAO data in the data set does not have substantial effects since the constraints from the BAO data on $\Lambda_{\rm s}$CDM are mostly due to the exclusion of the smaller $z_\dagger$ by the Galaxy BAO data as it was in Fig.~\ref{fig:evolution}. Although the Galaxy BAO data does not prefer the lowest $z_\dagger$ values for which the $\mathcal{D}(z)$ plot is substantially elevated, this effect cannot be rephrased as ``the larger the $z_\dagger$, the better agreement with the Galaxy BAO data" as we anticipated in the preliminary investigation in the previous section, because it seems from Fig.~\ref{fig:distance} that $\mathcal{D}(z)$ values that are slightly elevated compared to  $\Lambda$CDM would have better agreement with it. Indeed, including the BAO data in the analysis slightly elevates the plots of the $\Lambda$CDM model.
 
  \begin{figure}[t!]
    \centering
    \includegraphics[width=0.38\textwidth]{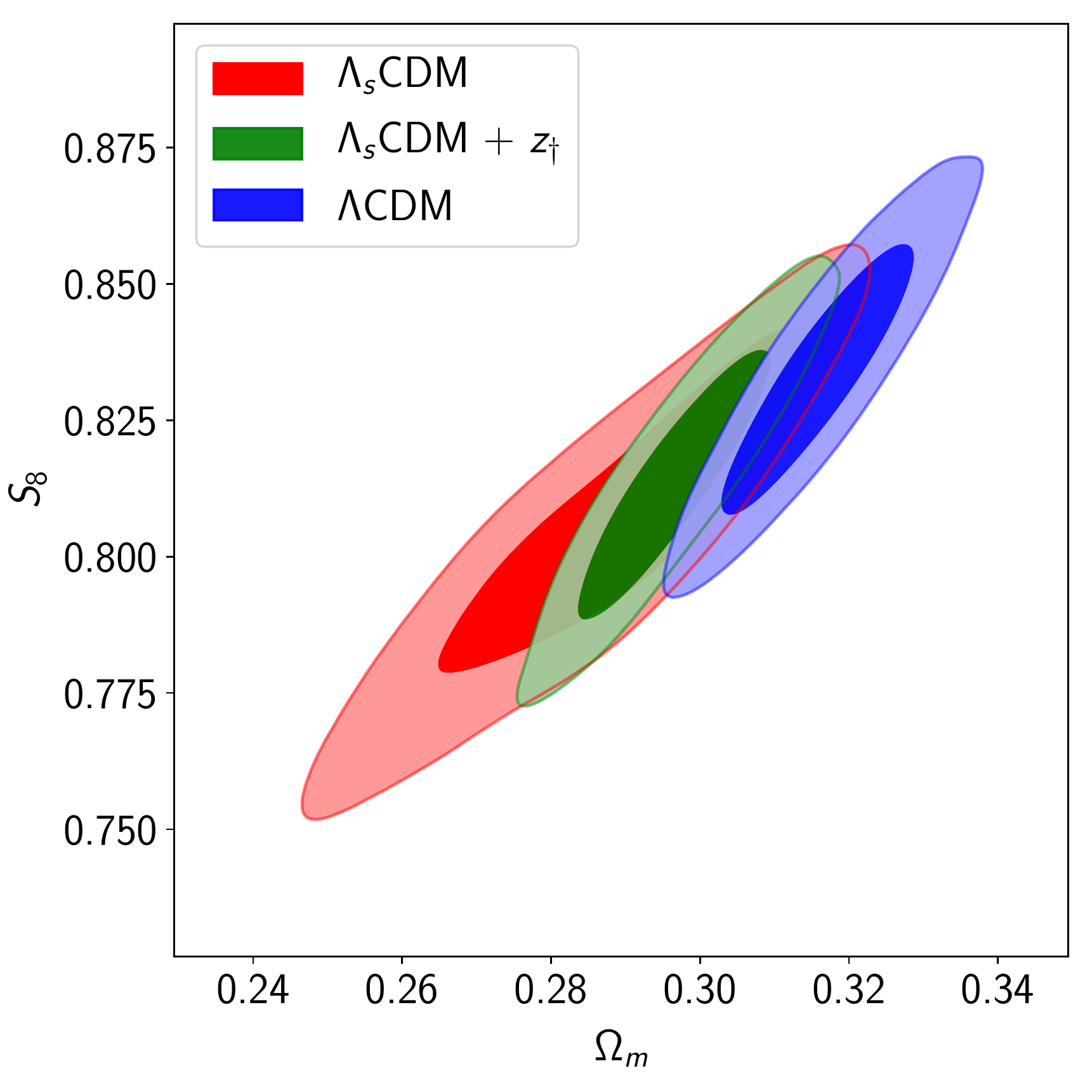}
    \caption{Two-dimensional (68\% and 95\% C.L.) marginalized distributions of $S_8$ versus $\Omega_{\rm m}$ from CMB data.
    }
    \label{fig:s8}
\end{figure}

 Table~\ref{tab:priors} also presents the values for the matter fluctuation amplitude parameter, $\sigma_8$. In the CMB-only analysis, the $\sigma_8$ value for the $\Lambda_{\rm s}$CDM model is slightly higher than that of the $\Lambda$CDM model. Including BAO data in our analysis increases the $\sigma_8$ value for $\Lambda_{\rm s}$CDM and decreases it for $\Lambda$CDM, resulting in an increased difference between the two models. It is important to include $\Omega_{\rm m}$ in the discussions of $\sigma_8$ since there is a discordance among various observational data in the $\sigma_8-\Omega_{\rm m}$ plane within $\Lambda$CDM that is usually quantified using $S_8$. Predictions of $S_8$ based on the CMB alone are in 2-3$\sigma$ tension with the measurements from dynamical low-redshift cosmological probes (weak lensing, cluster counts, redshift-space distortion) within the $\Lambda$CDM model. This is reflected in the CMB-only analysis in Table~\ref{tab:priors}, in which the value for $\Lambda$CDM reads $S_8=0.8332\pm0.0163$ compared to $S_8=0.766^{+0.020}_{-0.014}$ (KiDS-1000 weak lensing) \cite{Heymans:2020gsg}. 
Note that the measurement $S_8=0.804^{+0.032}_{-0.029}$ from the first-year data of HSC SSP~\cite{Hamana:2019etx} and also $S_8=0.800^{+0.029}_{-0.027}$ from KiDS-450+GAMA~\cite{vanUitert:2017ieu} remove this discrepancy; nonetheless, recent surveys still predict lower values, e.g., $S_8=0.776^{+0.017}_{-0.017}$ (DES weak lensing and galaxy clustering) \cite{DES:2021wwk}. 
Similar to the situation with the Ly-$\alpha$ measurements, alleviating the $S_8$ discrepancy within the $\Lambda$CDM model and its minimal extensions tends to exacerbate the $H_0$ tension \cite{DiValentino:2020vvd}; moreover, constraints on $S_8$ based on the Ly-$\alpha$ data are in agreement with the weak lensing surveys which probe similar late-time redshift scales as the Ly-$\alpha$ measurements \cite{Palanque-Delabrouille:2019iyz}.
So, it is conceivable that the $\Lambda_{\rm s}$CDM model provides a remedy for the $S_8$ discrepancy while retaining the better fit to the local measurements of $H_0$, as is the case for the Ly-$\alpha$ discrepancy. Indeed, Table~\ref{tab:priors} presents the $S_8$ values for the $\Lambda_{\rm s}$CDM models which are lower than those for $\Lambda$CDM in the CMB-only analysis, i.e., $S_8=0.8071\pm0.0210$ for the unrestricted and $S_8=0.8138\pm0.0166$ for the restricted model; see also Fig.~\ref{fig:s8}, which shows the $68\%$ and $95\%$ C.L. contours in the $S_8-\Omega_{\rm m}$ plane (notice that $\Lambda$CDM barely overlaps with $\Lambda_{\rm s}$CDM and does not overlap with $\Lambda_{\rm s}$CDM$+z_\dagger=2.32$ at $68\%$ C.L.). We see from the table that, although $\sigma_8$ is the smallest for the $\Lambda$CDM among the three models, its $\Omega_{\rm m}$ value greater than $0.3$ results in an increased $S_8$ value compared to its $\sigma_8$ value. In contrast, the $\Lambda_{\rm s}$CDM models have $\Omega_{\rm m}$ values lower than $0.3$ which result in decreased $S_8$ values compared to their $\sigma_8$ values. This results in the lower values of $S_8$ for $\Lambda_{\rm s}$CDM compared to the $\Lambda$CDM model. Note that lower $z_\dagger$ values correspond to lower $\Omega_{\rm m}$ values; this explains why the restricted $\Lambda_{\rm s}$CDM model exhibits a higher $S_8$ value. All three models have similar $S_8$ values when the BAO data is also included in the analysis; as before, this is due to the preference for higher $z_\dagger$ values by the Galaxy BAO data, since $\Lambda_{\rm s}$CDM approaches the $\Lambda$CDM model for larger $z_\dagger$ values and the $\Omega_{\rm m}$ values of the extended models are no longer less than 0.3. Thus, it appears that the $\Lambda_{\rm s}$CDM model partially reconciles the CMB data with the low redshift cosmological probes when $S_8$ is considered, and could potentially resolve the discrepancy in the absence of the Galaxy BAO data; however, for a robust conclusion on this matter, the constraints on $S_8$ from low redshift probes should also be investigated within the $\Lambda_{\rm s}$CDM model.

The constraints on the scalar spectral index $n_s$ do not differ substantially depending on the data sets and models. However, it is worth mentioning that $n_s$ in  $\Lambda$CDM is slightly smaller than the ones in the $\Lambda_{\rm s}$CDM models for the CMB-only analysis, while the situation is the opposite for the combined CMB+BAO analysis. We notice that, in $\Lambda$CDM, the inclusion of the BAO data decreases (increases) the marginalized value of $\omega_{\rm c }$ ($10^{2}\omega_{\rm b }$) obtained in the CMB-only analysis, and this effect is compensated by a shift in $n_s$ towards slightly larger values (see Ref.~\cite{Akrami:2018odb} for a similar result). Interestingly, it is the other way around and relatively more substantial for $\Lambda_{\rm s}$CDM: the inclusion of BAO data increases (decreases) the marginalized value of $\omega_{\rm c }$ ($10^{2}\omega_{\rm b }$) obtained in the CMB-only analysis, and this effect is compensated by a shift in $n_s$ towards smaller values.

We found no significant deviations in the constraints on the rest of the free parameters in Table~\ref{tab:priors}. $\theta_s$ is constrained robustly and almost the same in all cases, as expected.
$\tau_{\rm reio}$ and $\ln(10^{10}A_{\rm s})$ are almost the same for all three models in the CMB-only analysis. Including BAO in the analysis causes both $\tau_{\rm reio}$ and $\ln(10^{10}A_{\rm s})$ to go up, resulting in a slight decrease in the scaling of subhorizon anisotropies, i.e., in $A_{\rm s} e^{-2
\tau_{\rm reio}}$, for $\Lambda$CDM; in contrast, it causes both $\tau_{\rm reio}$ and $\ln(10^{10}A_{\rm s})$ to go down, resulting in a slight increase in the $A_{\rm s} e^{-2
\tau_{\rm reio}}$ value for the $\Lambda_{\rm s}$CDM models. This behavior of $\tau_{\rm reio}$ may be explained as follows. The reionization optical depth can be calculated using
$\tau_{\rm reio}=n_{\rm H}(0)c\sigma_{\rm T}\int_0^{z_{\rm max}}\dd{z} x_{\rm e}(z)\frac{(1+z)^2}{H(z)}$ (see, e.g., Ref.~\cite{Planck:2018vyg}), where $\sigma_{\rm T}$ is the Thomson scattering cross section, $n_{\rm H}(0)$ is the present-day total number of hydrogen nuclei, $x_{\rm e}(z)$ is the ratio of the number density at $z$ of the free electrons from reionization to the number of total hydrogen nuclei at $z$, and $z_{\rm max}$ is the integration bound that should be chosen high enough to allow the entire reionization to be captured (i.e., $z_{\rm max}\geq 50$). Although the shape of $x_{\rm e}(z)$ is not strictly constrained, it is expected to resemble a sigmoid function which is approximately zero for $z\geq10$ and slightly greater than unity for $z\leq 6$; it is modeled based on the hyperbolic tangent function by the \textit{Planck} Collaboration (2018) \cite{Planck:2018vyg}. Assuming $D_M(z_*)$ is the same for all three models in our analysis---which is closely related to the above integral---we expect lower $\tau_{\rm reio}$ values for the $\Lambda_{\rm s}$CDM models as a consequence of the suppression of the $z\gtrsim 10$ portion of the integral by $x_{\rm e}(z)$. 
This is because the $z\gtrsim 10$ portion constitutes a lower percentage of the total integral for $\Lambda$CDM compared to the other two since $H(z>z_\dagger)$ is greater for the $\Lambda$CDM model (so its contribution to the integral is smaller) in the CMB-only analysis. The results of the CMB-only analysis (see Table~\ref{tab:priors}) are in line with this argument. Following this logic, we expect the inclusion of the BAO data in the analysis to slightly increase $\tau_{\rm reio}$ for $\Lambda_{\rm s}$CDM for two reasons: first, the inclusion of the BAO data increases its $\omega_{\rm m}$ value, which implies a greater $r_*$ and hence greater $D_M(z_*)$ compared to the CMB-only analysis; second, this inclusion results in larger $z_\dagger$ values compared to the CMB-only analysis, making the model approach $\Lambda$CDM which we expect to have a higher $\tau_{\rm reio}$ value. Similar logic based on $\omega_{\rm m}$ (and $r_*$) may be used to expect a higher $\tau_{\rm reio}$ value for $\Lambda_{\rm s}$CDM$+z_\dagger=2.32$ and a lower value for $\Lambda$CDM. 
Surprisingly, the results in Table~\ref{tab:priors} are the opposite for all three models. This can be explained by changes in $n_{\rm H}(0)$ and $x_{\rm e}(z)$ with the inclusion of the BAO data, which are powerful enough to win over the effects explained above. Indeed, we see that the physical baryon density $\omega_{\rm b}$, which should naturally correlate positively with the total number of hydrogen nuclei $n_{\rm H}(0)$ and hence $\tau_{\rm reio}$, decreases for the $\Lambda_{\rm s}$CDM models and increases for $\Lambda$CDM.

Finally, to quantify which model performs better, we compute the Bayesian evidence used to perform a model comparison through the Jeffreys' scale \cite{Trotta:2008qt,Trotta:2017wnx}. In Table~\ref{tab:priors}, regarding the goodness of fit, we list $-2\ln{\mathcal{L}_{\rm max}}$ and the $\log$-Bayesian evidence ($\ln \mathcal{Z}$) for each of the models along with the Bayes' factor ($\Delta\ln \mathcal{Z}=\ln \mathcal{Z}_{\rm reference}-\ln \mathcal{Z}$)---the $\log$-Bayesian evidence for each of the models relative to the reference model, viz., the model with the lowest $|\ln \mathcal{Z}|$ value. The interpretation of the Bayes' factor according to the Jeffreys' scale is as follows: $0<\Delta\ln \mathcal{Z}\leq1$ implies that the strength of the evidence against the model compared to the reference model is weak/inconclusive, while the evidence is definite for $1\leq\Delta\ln \mathcal{Z}<3$, strong for $3\leq\Delta\ln \mathcal{Z}<5$, and very strong for $\Delta\ln \mathcal{Z}>5$ \cite{Jeffreys}. We see from Table~\ref{tab:priors} that all the models fit equally well to the data for both the CMB-only and the combined CMB+BAO analyses. For the CMB-only analysis, the restricted $\Lambda_{\rm s}$CDM model is the reference model and there is weak evidence against the other two models. In the case of the combined CMB+BAO data analysis, $\Lambda$CDM is the reference model, and the unrestricted $\Lambda_{\rm s}$CDM model departs from it with definite evidence due to the presence of the additional free parameter $z_\dagger$. However, we note that $\Lambda_{\rm s}$CDM agrees better with the model-independent measurements of $H_0$ and $M_B$, the constraints on $\omega_{\rm b}$ from BBN, and the constraints on $S_8$ from low-redshift probes, which are excluded in the observational analyses in the current work.

\section{Conclusions}
\label{sec:conc}

In this paper, we first discussed the possibility that dark energy models with energy densities that attain negative values in the past can alleviate the $H_0$ tension, as well as the discrepancy with the Ly-$\alpha$ BAO measurements, both of which prevail within the $\Lambda$CDM model. The so-called graduated dark energy \cite{Akarsu:2019hmw}, having this feature, when restricted to its parameter space constrained by observations, is phenomenologically well approximated by a cosmological constant which switches sign at redshift $z\approx2.32$ to become positive today. It, however, accommodates the weak energy condition and the bounds on the speed of sound at its limit of cosmological constant, which comes with a sign-switching feature in contrast to the usual cosmological constant ($\Lambda$). This led the authors of Ref.~\cite{Akarsu:2019hmw} to conjecture that the Universe transitioned from AdS vacua to dS vacua at $z\approx2.3$. Inspired by this, we have introduced the $\Lambda_{\rm s}$CDM model, which promotes the usual cosmological constant assumption of the standard $\Lambda$CDM model to a sign-switching cosmological constant ($\Lambda_{\rm s}$).

The $\Lambda_{\rm s}$CDM model, neglecting radiation, corresponds to gluing two Friedmann-Lema\^{i}tre models at $z=z_\dagger$: one with a cosmological constant that yields a negative value of $\Lambda=-\Lambda_{\rm s0}<0$, which is superseded by the other with a cosmological constant that yields a positive value of $\Lambda=\Lambda_{\rm s0}>0$.
The deviation of this model from $\Lambda$CDM is controlled by its only additional parameter $z_\dagger$, the redshift at which the cosmological constant switches sign, for which the limit $z_\dagger\to\infty$ gives the $\Lambda$CDM model. Before directly confronting the model with observational data, we carried out a preliminary investigation to assess the reasonable range of $z_\dagger$, and its effects on the dynamics of the Universe. We fixed the physical matter density at the CMB last scattering and the comoving angular diameter distance to last
scattering to those of $\Lambda$CDM to ensure good consistency with the CMB data. We then found that $H_0$ is inversely correlated with $z_\dagger$, and for $z_\dagger=1.5$ it reaches $\approx74.5~{\rm km\, s^{-1}\, Mpc^{-1}}$. 
It is comforting that this value is already consistent with even the highest values of model-independent local measurements of $H_0$ by the SH0ES Collaboration, because the values of $z_\dagger$ less than about $1.5$ are strongly disfavored by the Galaxy BAO measurements. Next, we showed that, unlike many other models with late-time modifications to $\Lambda$CDM suggested to address the $H_0$ tension, the $\Lambda_{\rm s}$CDM model respects the internal consistency of the methodology used by the SH0ES Collaboration to estimate $H_0$ and $M_B$ (SnIa absolute magnitude), and therefore, within the $\Lambda_{\rm s}$CDM model, the amelioration of the SH0ES $H_0$ tension should be accompanied by an amelioration of the $M_B$ tension. 
Also, it is interesting to observe that, as long as $z_\dagger\lesssim2.34$, the model remains in excellent agreement with the Ly-$\alpha$ measurements even for $z_\dagger\sim 1.1$, which barely satisfies the condition that we live in an ever-expanding Universe; a good agreement with the Ly-$\alpha$ data is an intrinsic feature of the $\Lambda_{\rm s}$CDM model as long as $z_\dagger\lesssim2.34$. In light of these discussions, we came to the conclusion that the Galaxy and Ly-$\alpha$ BAO measurements would determine the lower and upper bounds of $z_\dagger$, respectively. We leave the interesting possibility of violating the condition $z_\dagger\gtrsim 1.1$ to future works. In this case, the Universe passes through a contraction phase, which in turn breaks the one-to-one correspondence between cosmic time and redshift, resulting in signals from the same redshift but two different ages of the Universe.

We carried out a robust observational analysis first with the full CMB data, and then with the combined CMB+BAO data set, to constrain the parameters of the $\Lambda_{\rm s}$CDM model, its particular case having $z_\dagger=2.32$, and the $\Lambda$CDM model. We found that the CMB data alone do not constrain $z_\dagger$, but the combined CMB+BAO data set predicts $z_\dagger=2.44\pm0.29$ (68\% C.L.) with a peak at $z_\dagger\approx2.33$ in the posterior. We found slightly positive evidence (Bayesian) in favor of $\Lambda$CDM over the $\Lambda_{\rm s}$CDM model for the CMB+BAO data set, while all of the models fit the data equally well. 
However, the $\Lambda_{\rm s}$CDM model still stands in a privileged position as it removes the discrepancy with the Ly-$\alpha$ measurements, has better agreement with the BBN constraints on the physical baryon density ($\omega_{\rm b}$), provides a lower $S_8$ value based on the \textit{Planck} data which alleviates its discordance with some low-redshift cosmological probes, predicts a higher absolute magnitude $M_B$ value for SnIa which is in better agreement with its locally determined constraints obtained by Cepheid calibrators, and also alleviates the $H_0$ tension, especially when the TRGB $H_0$ measurement is considered. Also, it is important to note that the amelioration in the last four is not captured by the Bayesian evidence as the data/priors on $\omega_{\rm b}$ from BBN, on $H_0$ from local measurements, on $S_8$ from dynamical cosmological probes (weak lensing, cluster counts, redshift-space distortion), and on $M_B$ from its local determinations obtained by Cepheid calibrators are not included in our observational analyses.
These improvements come at the cost of worsening in describing the comoving angular diameter distance measurements from the Galaxy BAO; in fact, the preference of larger $z_{\dagger}$ values by the Galaxy BAO data prevents the $\Lambda_{\rm s}$CDM model from reaching its full potential of having an excellent agreement with even the highest local $H_0$ measurements in consistency with the constraints on $M_B$ from Cepheid calibrators, and the lowest $S_8$ measurements. In this regard, when BAO data is considered, the $\Lambda$CDM model is in conflict with the Ly-$\alpha$ measurements, while the $\Lambda_{\rm s}$CDM model is in conflict with the Galaxy BAO measurements; forthcoming observations will be crucial in determining which model is preferred by nature. 
However, there is an asymmetry between the two models in the sense that, if new observations are able to remove the conflict of $\Lambda$CDM  with the Ly-$\alpha$ measurements, the discrepancy with the BBN constraints on $\omega_{\rm b}$, the $S_8$ discrepancy, and the unnerving $H_0$  and $M_B$ tensions remain; in contrast, if new observations are able to remove the conflict of $\Lambda_{\rm s}$CDM with the Galaxy BAO measurements, it can work even better in alleviating the $H_0$  and $M_B$ tensions while retaining its superior agreement with the BBN constraints on $\omega_{\rm b}$, the Ly-$\alpha$ measurements, and the constraints on $S_8$ from dynamical probes. Confronting the $\Lambda_{\rm s}$CDM model by considering BBN and/or $M_B$ priors and additional observational data from weak lensing, cluster counts, SnIa, cosmic chronometers etc., along with the CMB and BAO data used in this study, would allow a more extensive evaluation of the model, and a better assessment of the importance of the Galaxy BAO data with regard to the $\Lambda_{\rm s}$CDM model.

The assumptions of the $\Lambda_{\rm s}$CDM model---that the sign transition of $\Lambda_{\rm s}$ happens instantaneously and that the value of $\Lambda_{\rm s}$ is exactly the opposite of its present-day value before the transition---might be too restrictive both phenomenologically and (bearing in mind that such phenomena should eventually be realized via a mechanism from fundamental theories of physics) theoretically. Accordingly, it is possible to think of two natural phenomenological extensions to the $\Lambda_{\rm s}$CDM model: first, the sign-switching cosmological constant described here by a step function can be extended via smooth sigmoid functions so that the rapidity of the switch can also be controlled; second, one can consider a scenario in which the cosmological constant reaches its present-day positive value by an arbitrary shift in its value rather than a sign switch, and constrain the amount of change in its value as an extra parameter (in this case, additional scenarios with a vanishing or positive-valued cosmological constant in the past are also possible, and the shift in the cosmological constant need not be positive, but obviously a negative shift is not expected considering what we have learned from this current study); a third model can be constructed by combining these two, which would be the most natural one. From the perspective of theoretical physics constructions that would underlie the sign switch (or the value transition) feature, these extensions will be more reasonable and expand the space of possible theoretical mechanisms. 

One such mechanism can be straightforwardly realized in unimodular gravity (UG)  \cite{Ellis:2010uc,Josset:2016vrq} if the usual vacuum energy of QFT suddenly or gradually diffuses to the cosmological constant (which could be negative in the past) and uplifts it to its present-day observed value. Since the usual vacuum energy of QFT does not gravitate in UG, the change in its energy density has no effect on the dynamics of the Universe, but the change in the value of the cosmological constant (which arises naturally as an integration constant in UG and contributes to the field equations as a geometrical component) does affect the dynamics; thus, this mechanism can produce the exact phenomenology of $\Lambda_{\rm s}$CDM and all three of its extensions depending on the functional form and amount of the diffusion. Recently, such a mechanism within UG---for which the diffusion, instead of happening from the usual vacuum energy of the QFT, happens from the matter sector to the cosmological constant---was studied both theoretically and phenomenologically to address the $H_0$ tension \cite{Perez:2017krv, Perez:2018wlo, Perez:2019gyd,Perez:2020cwa,LinaresCedeno:2020uxx,preprint}; however, note that this scenario is different from $\Lambda_{\rm s}$CDM and its above-mentioned extensions, as this mechanism uses some energy budget from the energy density of the matter sector. The sign switch feature of the $\Lambda_{\rm s}$CDM model is reminiscent of the so-called Everpresent $\Lambda$ model \cite{Ahmed:2002mj,Zwane:2017xbg}, which was suggested for addressing the $H_0$ and Ly-$\alpha$ tensions, in which the observed $\Lambda$ fluctuates between positive and negative values with a magnitude comparable to the cosmological critical energy density about a vanishing mean, $\braket{\Lambda}=0$, in any epoch of the Universe, in accordance with a long-standing heuristic prediction of the causal set approach to quantum gravity \cite{Surya:2019ndm}. Nevertheless, the $\Lambda_{\rm s}$CDM model suggests that the sign switch of the cosmological constant is a single event that happens in the late Universe at $z\sim2$. If we stick to this, namely, a very rapid single transition or its limiting case a single instantaneous (discontinuous) transition in the value of the cosmological constant, then it would be more reasonable to look for a potential origin of this phenomenon in a theory of fundamental physics by considering it as a first-order phase transition. See Ref.~\cite{Mazumdar:2018dfl} for a recent review on well-known cosmic phase transitions. Recently, the phase transition approach has been used to address the $H_0$ tension; see, e.g., Refs.~\cite{Banihashemi:2018oxo,Banihashemi:2018has,Banihashemi:2020wtb}, which considered that the DE density behaves like the magnetization of the Ising model and presented a realization of this behavior within the Ginzburg-Landau framework---which is an effective field theory (EFT) describing the physics of phase transitions without any dependence on the details of relevant microstructures---and Ref.~\cite{Farhang:2020sij}, which considered a gravitational phase transition that is justified from the EFT point of view. The model studied in Ref.~\cite{Banihashemi:2018oxo} is phenomenologically similar but not equivalent to the one-parameter extension of $\Lambda_{\rm s}$CDM with an arbitrary shift in the value of the cosmological constant, as (in contrast to our approach in this work) the cosmological constant is not allowed to take negative values (and thereby the model addresses the $H_0$ tension with a shift in the value of the cosmological constant at very low redshifts, viz., $z_{\rm t}=0.092^{+0.009}_{-0.062}$, signaling that it could suffer from the $M_B$ tension; see Sec.~\ref{sec:zdagger}). Given the promising advantages of having a negative cosmological constant for $z\gtrsim2$ regarding the cosmological tensions, as discussed in this work, and that negative cosmological constant is a theoretical sweet spot---AdS space/vacuum is welcome due to the AdS/CFT correspondence \cite{Maldacena:1997re} and is preferred by string theory and string theory motivated supergravities \cite{Bousso:2000xa}---it would be most natural to associate this phenomena with a possible phase transition from AdS to dS that is derived in string theory, string theory motivated supergravities, and theories that find motivation from them. The phase transitions from AdS to dS (most compatible with our approach and findings), Minkowski (corresponding to $\Lambda=0$) to dS, and dS to dS pertain to active area of research in theoretical physics, but finding four-dimensional dS spacetime solutions has been a vexing quest and so far the AdS to dS transition has rarely been directly linked to physical cosmology and particularly dark energy in the literature, see, e.g., Ref.~\cite{Sato:1980yn,Caldwell:2005xb,Polchinski:2006gy,Gupt:2013poa,Samart:2020klx,DeAlwis:2019rxg,Samart:2020mnn,Samart:2020qya,Camanho:2015zqa,Cadoni:2013gza,Kim:2007ix,Berglund:2021xlm,Berglund:2019ctg,Heckman:2018mxl,Heckman:2019dsj,Berglund:2019pxr,Banerjee:2019fzz,Cicoli:2018kdo,Kallosh:2001gr,Kallosh:2002gg,Banerjee:2018qey,Akrami:2017cir,Dutta:2021bih,Franchino-Vinas:2019nqy,Garriga:2002tq,Garriga:2003hj,Biswas:2009fv,Prokopec:2011ce,Dong:2011uf,Camanho:2013uda,Denef:2018etk,Kachru:2018aqn,Kiritsis:2019wyk}.

Finally, both of the above-mentioned extensions of $\Lambda_{\rm s}$CDM introduce two extra free parameters on top of $\Lambda$CDM, and their combination introduces three. Despite their excess number of free parameters (subject to observational constraints), both the promising features of the $\Lambda_{\rm s}$CDM model, and the fact that these phenomenological models could act as a guide and a cosmological testing ground for the fundamental physics theories giving rise to their phenomena, suggest that these extensions are worth further studying. Regarding the rapidity of the AdS to dS transition in a string theory setup, note the comments against continuous variation of the cosmological constant, which could necessitate an instantaneous transition as in $\Lambda_{\rm s}$CDM \cite{Polchinski:2006gy}. In this sense, a two-parameter extension of $\Lambda$CDM with an instantaneous arbitrary shift in the value of the cosmological constant could be the most natural next phenomenological step of our work presented in this paper.

\begin{acknowledgments}
The authors are grateful to the referee for valuable comments and suggestions. The authors thank Mehmet \"Ozkan for discussions. \"{O}.A. acknowledges the support by the Turkish Academy of Sciences in the scheme of the Outstanding Young Scientist Award  (T\"{U}BA-GEB\.{I}P). E.\"{O}.~acknowledges the support by The Scientific and Technological Research Council of Turkey (T\"{U}B\.{I}TAK) in scheme of 2211/A National PhD Scholarship Program. J.A.V. acknowledges the support provided by FOSEC SEP-CONACYT Investigaci\'on B\'asica A1-S-21925, PRONACES-CONACYT/304001/2020, and UNAM-DGAPA-PAPIIT IA104221.
\end{acknowledgments}

\end{document}